\documentclass[11pt]{article}
\RequirePackage{xspace}
\usepackage{a4}
\usepackage{latexsym} 
\usepackage{amssymb}  
\usepackage{amsfonts}  
\usepackage{graphicx}
\usepackage{epsfig}
\usepackage{pslatex}
\usepackage{hyperref}
\usepackage{relsize}
\usepackage{amsmath}
\usepackage[english]{babel}
\usepackage{cite}
\usepackage{xspace}

\def\auth#1{~\hfill {\it #1}~\\}
\def\spk#1#2{#1\, ({\it #2})}

\def\tit#1{~\\\noindent{\bf #1}}

\newenvironment{wparticipants}{\noindent {\sf PARTICIPANTS:}\\[2mm]}{\\[5mm]}

\newfont{\Thuge}{cmr10 scaled 8000}
\newfont{\thuge}{cmr10 scaled 6000}

\newcommand{\sqrtsnn}{\sqrt{s_{_{NN}}}}

\def\mean#1{\ensuremath{\left<#1\right>}}

\newcommand{\SSpt}{$p_T$\xspace}
\newcommand{\SSsqrtsL}{$\sqrt{s}=$~900~GeV\xspace}
\newcommand{\SSsqrtsH}{$\sqrt{s}=$~7~TeV\xspace}
\newcommand{\SSsqrtsHs}{7~TeV\xspace}

\newcommand \Pom {I\!\!P}

\def\ttt#1{\texttt{\small #1}}

\def\beq{\begin{eqnarray}}
\def\eeq{\end{eqnarray}}

\newcommand{\phojet}{{\sc phojet}}
\newcommand{\dpmjet}{{\sc dpmjet}}
\newcommand{\epos}{{\sc epos}}

\newcommand{\qgsjet}{{\sc qgsjet}} 
\newcommand{\sibyll}{{\sc sibyll}}

\newcommand{\pythia}{{\sc pythia}}
\newcommand{\herwig}{{\sc herwig}}
\newcommand{\sherpa}{{\sc sherpa}}

\newcommand{\fluka}{{\sc fluka}}
\newcommand{\corsika}{{\sc corsika}}

\newcounter{zyxabstract}     

\def\pbar       {\kern 0.2em\overline{\kern -0.2em p}{}\xspace}

\def\lambdacbar {\kern 0.2em\overline{\kern -0.2em \Lambda}{}_c^-\xspace}

\def\Bbar    {\kern 0.18em\overline{\kern -0.18em B}{}\xspace}

\def\Kbar  {\kern 0.2em\overline{\kern -0.2em K}{}\xspace}

\mathchardef\Upsilon="7107

\newcommand{\cms}{\sqrt{s}}
\newcommand{\expval}[1]{\langle #1 \rangle}
\newcommand{\nch}{N_{\rm ch}}

\newcommand{\as}{\alpha_s}
\newcommand{\Lb}{\left(}
\newcommand{\Rb}{\right)}

\begin{document}

\begin{center}
{\Large \bf Hadron-Hadron and Cosmic-Ray Interactions at multi-TeV Energies} \\ 
~\\
{\Large Mini-proceedings ECT* Workshop, Trento,  Nov. 28 - Dec. 3, 2010}\\[0.3cm]

B.~Alessandro$^{1}$, 
D.~Bergman$^{2}$,
M.~Bongi$^{3}$, 
A.~Bunyatyan$^{4}$,
L.~Cazon$^{5}$, 
D.~d'Enterria~$^{6,7}$,
I.~de~Mitri$^{8,9}$, 
P.~Doll$^{10}$, 
R.~Engel$^{11}$,
K.~Eggert$^{7,12}$, 
M.~Garzelli$^{13,14}$, 
L.~Gerhardt$^{15}$, 
S.~Gieseke$^{16}$, 
R.~Godbole$^{7}$, 
J.F.~Grosse-Oetringhaus$^{7}$, 
G.~Gustafson$^{17}$, 
T.~Hebbeker$^{18}$, 
L.~Kheyn$^{19}$, 
J.~Kiryluk$^{15}$, 
P.~Lipari$^{20}$, 
S.~Ostapchenko$^{21}$, 
T.~Pierog$^{11}$, 
O.~Piskounova$^{22}$, 
J.~Ranft$^{23}$, 
A.~Rezaeian$^{24}$, 
A.~Rostovtsev$^{25}$,
N.~Sakurai$^{26}$,
S.~Sapeta$^{27}$, 
S.~Schleich$^{28}$, 
H.~Schulz$^{29}$, 
T.~Sj\"ostrand$^{17}$, 
L.~Sonnenschein$^{18}$,
M.~Sutton$^{30}$,
R.~Ulrich$^{31}$, 
K.~Werner$^{32}$, and
K.~Zapp$^{33}$\\[0.5cm]

{\small

{\it $^{1}$ INFN Sezione di Torino, Torino, Italy}\\
{\it $^{2}$ University of Utah, Dept. Phys. \& Astron., Salt Lake City, UT 84112, USA}\\
{\it $^{3}$ INFN Sezione di Firenze, Via Sansone 1, I-50019 Sesto Fiorentino, Firenze, Italy}\\
{\it $^{4}$ DESY, Notkestrasse 85, 22607 Hamburg, Germany}\\
{\it $^{5}$ LIP, Av. Elias Garcia 14 - 1o, 1000-149 Lisbon, Portugal}\\
{\it $^{6}$ ICREA \& ICC-UB, Universitat de Barcelona, 08028 Barcelona, Catalonia}\\
{\it $^{7}$ CERN, PH Department, CH-1211 Geneva 23, Switzerland}\\
{\it $^{8}$ Dipartimento di Fisica, Universit\`a del Salento, I-73100, Lecce, Italy}\\
{\it $^{9}$ INFN Sezione di Lecce, I-73100, Lecce, Italy}\\
{\it $^{10}$ Institut f\"ur Experimentelle Kernphysik, KIT Campus Sued, 76021 Karlsruhe, Germany}\\
{\it $^{11}$ Karlsruhe Institute of Technology, P.O.\ Box 3640, 76021 Karlsruhe, Germany}\\
{\it $^{12}$ Case Western Reserve University, Cleveland, USA}\\
{\it $^{13}$ INFN Sezione di Milano, Milano, Italy}\\
{\it $^{14}$ Depto. F\'{i}sica Te\'orica y del Cosmos y CAFPE, Univ. de Granada, 18071 Granada, Spain}\\
{\it $^{15}$ Lawrence Berkeley National Laboratory, 1 Cyclotron Road, 94720 Berkeley, USA}\\
{\it $^{16}$ Institute for Theoretical Physics, Karlsruhe Institute of Technology, 76128 Karlsruhe, Germany}\\
{\it $^{17}$ Dept. of Astronomy and Theoretical Physics, Lund University, SE-223 62 Lund, Sweden}\\
{\it $^{18}$ RWTH Aachen University, Phys. Inst. IIIA, 52056 Aachen, Germany}\\
{\it $^{19}$ D.V. Skobeltsyn Institute of Nuclear Physics, Moscow State University, 119992 Moscow, Russia}\\
{\it $^{20}$ INFN Sezione di Roma, Roma, Italy}\\
{\it $^{21}$ NTNU, Institutt for Fysikk, 7491 Trondheim, Norway}\\
{\it $^{22}$ P.N.Lebedev Physical Institute of Russian Academy of Science, Moscow, Russia}\\
{\it $^{23}$ Siegen University, Siegen, Germany}\\
{\it $^{24}$ Depto. F\'\i sica, Univ. T\'ecnica Federico Sta. Ma., Casilla 110-V, Valparaiso, Chile}\\
{\it $^{25}$ Institute for Theor. \& Exp. Physics, B.Cheremushkinskaya 25, 117218 Moscow, Russia}\\
{\it $^{26}$ Faculty of Science, Osaka City Univ., 3-3-138 Sugimoto, Sumiyoshi-ku, Osaka 558-8585, Japan}\\
{\it $^{27}$ LPTHE, UPMC Univ.~Paris 6 and CNRS UMR 7589, Paris, France}\\
{\it $^{28}$ TU Dortmund, Univ. Dortmund, Exp. Physik 5, Otto-Hahn-Strasse 4, Dortmund, Germany}\\
{\it $^{29}$ Humboldt University, Inst. f. Physik, Newtonstr. 15, 12489 Berlin, Germany}\\
{\it $^{30}$ Dept. of Physics and Astronomy, University of Sheffield, S3 7RH Sheffield, UK}\\
{\it $^{31}$ Dept. of Physics, 104 Davey Lab, Penn State University, 16802 University Park, USA}\\
{\it $^{32}$ SUBATECH, 4 rue Alfred Kastler, BP 20722, 44307 Nantes Cedex 3, France}\\
{\it $^{33}$ IPPP, 
Science Laboratories, Durham Univ., DH1 3LE Durham, UK}\\
}

\vspace{0.4cm}
{\large ABSTRACT}
\end{center}
The workshop  on ``Hadron-Hadron and Cosmic-Ray Interactions at multi-TeV Energies'' 
held at the ECT* centre (Trento) in Nov.-Dec. 2010 gathered together
both theorists and experimentalists to discuss issues of the physics of 
high-energy hadronic interactions of common interest for the particle, nuclear 
and cosmic-ray communities. 
QCD results from collider experiments -- mostly from the LHC but also
from the Tevatron, RHIC and HERA -- were discussed and compared to various
hadronic Monte Carlo generators, aiming at an improvement of our theoretical
understanding of soft, semi-hard and hard parton dynamics.
The latest cosmic-ray results from various ground-based observatories were 
also presented with an emphasis on the phenomenological modeling of the first hadronic
interactions of the extended air-showers generated in the Earth atmosphere.
These mini-proceedings consist of an introduction and short summaries of 
the talks presented at the meeting.



\tableofcontents


\section*{Introduction}

\noindent
The origin and nature of cosmic rays (CRs) with energies between $10^{15}$~eV
and the so-called Greisen-Zatsepin-Kuzmin (GZK) cut-off at about $10^{20}$\,eV~\cite{gzk}, recently measured by 
the HiRes~\cite{Abbasi-2007-PRL-100-101101} and Auger~\cite{Abraham:2008ru} experiments, 
remains a central open question in high-energy astrophysics
with very interesting connections to particle physics and, in particular, to Quantum-Chromo-Dynamics 
(QCD) at the highest energies ever studied. One key to solving this question is the determination 
of the elemental composition of cosmic rays in this energy range. The candidate particles, ranging 
from protons to nuclei as massive as iron, generate ``extensive air-showers'' (EAS) in 
interactions with air nuclei when entering the Earth's atmosphere.
The determination of the primary energy  and mass relies on hadronic Monte Carlo (MC) models 
which describe the interactions of the primary cosmic-ray in the upper atmosphere.\\

\noindent
The bulk of particle production in such high-energy hadronic collisions can still not 
be calculated within first-principles QCD and general principles such as unitarity and 
analyticity (as implemented in Regge-Gribov theory) are often combined with perturbative 
QCD predictions for high-$p_{T}$ processes, constrained by the existing collider data 
($E_{lab}\lesssim$~10$^{15}$~eV). Important theoretical issues at these energies are the
understanding of diffractive and elastic hadronic scattering contributions, the description of hadronic forward 
fragmentation and multi-parton interactions (``underlying event''), and the effect of high parton densities
(``gluon saturation'') at small values of parton fractional momentum $x=p_{\mbox{\tiny{\it parton}}}/p_{\mbox{\tiny{\it proton}}}$.
Indeed, at these energies, the relevant Bjorken-$x$ values are as low as 10$^{-7}$, where effects 
like gluon saturation and multi-parton interactions, particularly enhanced with nuclear targets, 
are expected to dominate the early hadron collision dynamics.\\

\noindent
The current energy frontier for hadron collisions in the laboratory is reached at the 
Large Hadron Collider (LHC), currently under operation at CERN. The measurement of inclusive 
hadron production observables in proton-proton, proton-nucleus, and nucleus-nucleus collisions, at 
LHC energies (equivalent to $E_{lab} \approx 10^{17}$~eV) will provide very valuable information
on high-energy multiparticle production, and allow for more reliable determinations of the CR energy 
and composition around the GZK cutoff. In the high luminosity phase of LHC, each bunch crossing will lead to several 
proton-proton interactions, increasing even more the importance of understanding the background 
from diffractive and soft particle production. Semi-hard particle physics will allow one
to test the boundaries of the applicability of perturbative QCD in the region where low-$x$ gluon 
saturation phenomena become increasingly important and may even dominate 
particle production.\\

\noindent
All LHC experiments feature detection capabilities with a wide phase-space coverage without parallel, 
in particular in the forward direction, compared to previous colliders~\cite{d'Enterria:2007dt}. 
Such capabilities allow for a (fast) measurement of global hadron-hadron collision properties 
(inelastic -- including diffractive -- cross sections, particle multiplicity and energy flows 
as a function of $p_T$ and pseudorapidity, ...) even with the moderate statistics 
of a first $p p$ and $Pb Pb$ run.\\

\noindent
The aim of the Workshop was to discuss theoretical and experimental
issues connected to hadronic interactions of common interest for high-energy 
particle and cosmic-ray physics. With the recent high-quality
cosmic-ray results from the HiRes and Auger experiments and the
first available LHC data it seemed a timely
moment to have such a meeting in autumn 2010. The Workshop brought 
together experts, both theorists and experimentalists, in QCD and
cosmic-ray physics in view of expanding the mutually beneficial
interface between two communities currently exploring the physics of
strong interactions at the highest energies accessible.
\noindent The talks and discussions on various topics:
\begin{itemize} 
\item QCD predictions for high-energy multiparticle production 
and their implementation in hadronic Monte Carlo generators: \pythia, \herwig, \sherpa, \phojet, \dpmjet,
  \qgsjet, \sibyll, \epos, {\sc qgsm}, \fluka;
\item theoretical and experimental developments on diffractive and elastic scattering at high energies;
\item theoretical approaches of multi-parton dynamics and underlying event in hadronic collisions;
\item theoretical and experimental developments on low-$x$ QCD and inclusive particle production;
\item theoretical developments on modeling of cosmic-ray showers;
\item latest experimental QCD results at colliders: LHC (ATLAS, CMS, ALICE, LHCb, TOTEM, LHCf), 
Tevatron, RHIC, and HERA;
\item latest cosmic-rays measurements in the $10^{15}$-$10^{20}$\,eV range:
Auger, HiRes, TA, Kascade-Grande, Argo, IceCube;
\end{itemize}
were organized around four main blocks:
\begin{enumerate} 
\item Hadronic collisions at multi-TeV energies: Experimental results
\item Hadronic collisions at multi-TeV energies: Theory
\item Cosmic-rays at Ultra-High Energies: Experimental results
\item Cosmic-rays at Ultra-High Energies: Theory
\end{enumerate}

\noindent
These mini-proceedings include a short summary of each talk including 
relevant references, the list of participants and the workshop programme. We felt that such 
a format was more appropriate than full-fledged proceedings. Most results are or 
will soon be published and available on arXiv. Most of the talks can also be downloaded 
from the workshop website:\\

\centerline{ \texttt{http://www.cern.ch/CRLHC10/}}

\bigskip

\noindent
Support of the European Community Research Infrastructure Action under
the FP7 'Capacities Specific Programme' is acknowledged.
We thank the ECT* management and secretariat, in particular Cristina Costa, for the helpful 
cooperation prior and during the workshop and all participants for their valuable contributions. We believe 
that this was only the first workshop of this kind and look forward to similar meetings in the future.\\

\vspace{0.6cm}

\noindent
{\sc David d'Enterria, Ralph Engel, Torbj\"orn Sj\"ostrand}

\newpage

\section{Hadronic collisions at multi-TeV energies: Experiments}

\tit{First QCD results from the ATLAS Collaboration}

\auth{Mark Sutton\footnote{On behalf of the ATLAS collaboration.} (University of Sheffield)}

Since the first LHC operation in Nov. 2009 the ATLAS experiment 
has collected data at several proton-proton centre-of-mass energies 
with an integrated luminosity of 45~pb$^{-1}$ in collisions at 7~TeV.  
This large sample has enabled many QCD-related analyses to be performed 
spanning the entire kinematic range from soft QCD in minimum
bias interactions~\cite{atlasone,atlasthree} 
through to the study 
of the underlying event~\cite{atlasfour} in events with at least 
one energetic track, through to production of jets with high 
transverse momentum $p_T$~\cite{atlasfive,atlassix}. 
Events with a prompt photon or a $Z$ or $W$ have also been observed 
copiously~\cite{atlasseven,atlaseight} as have events 
where the gauge boson is produced in conjunction with a high-$p_T$ jet~\cite{atlasnine}.

\begin{figure}[htp]
\begin{minipage}{17cm}
\includegraphics[height=7.3cm,width=7.5cm]{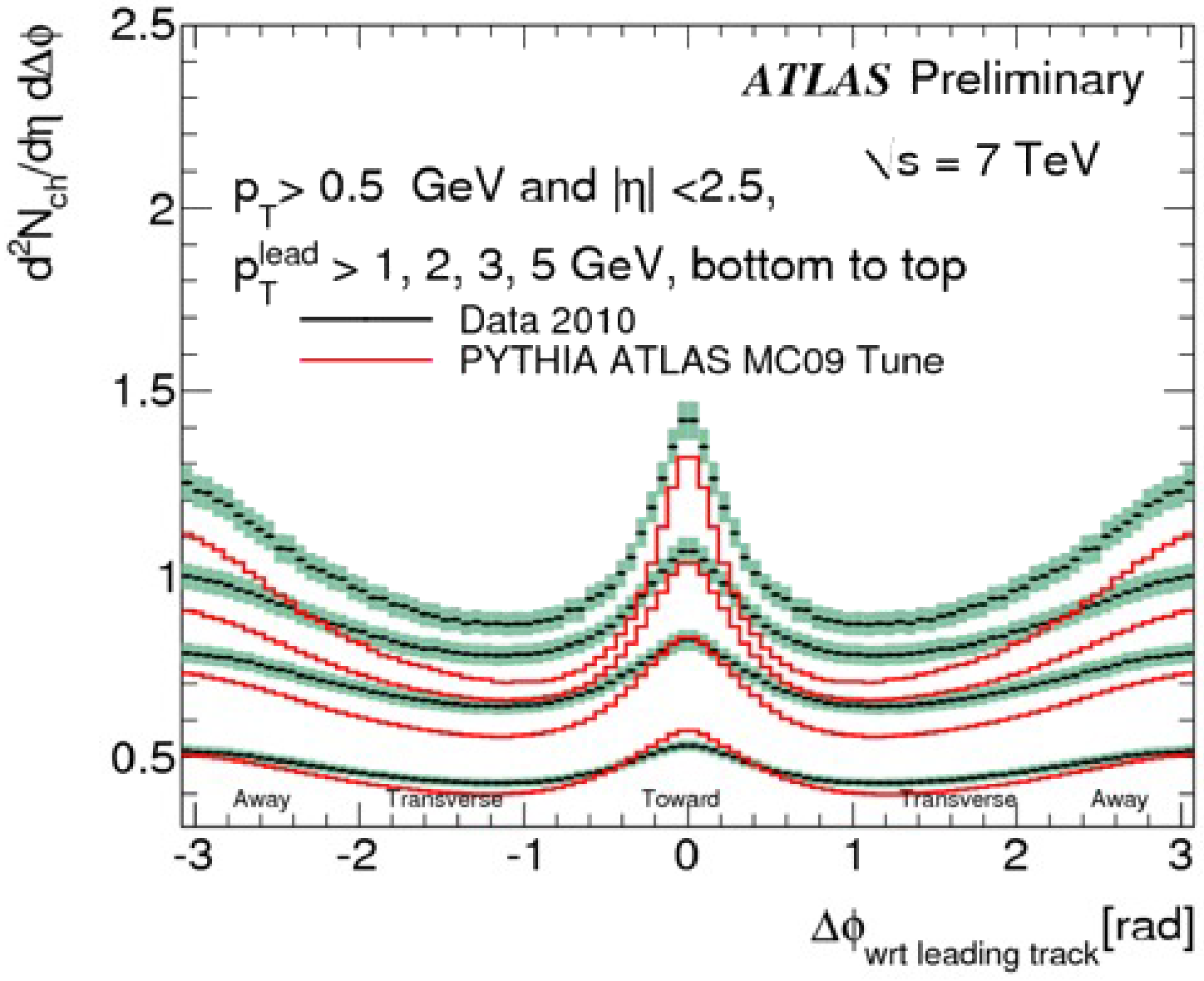}
\hspace{2mm}
\includegraphics[width=7.8cm]{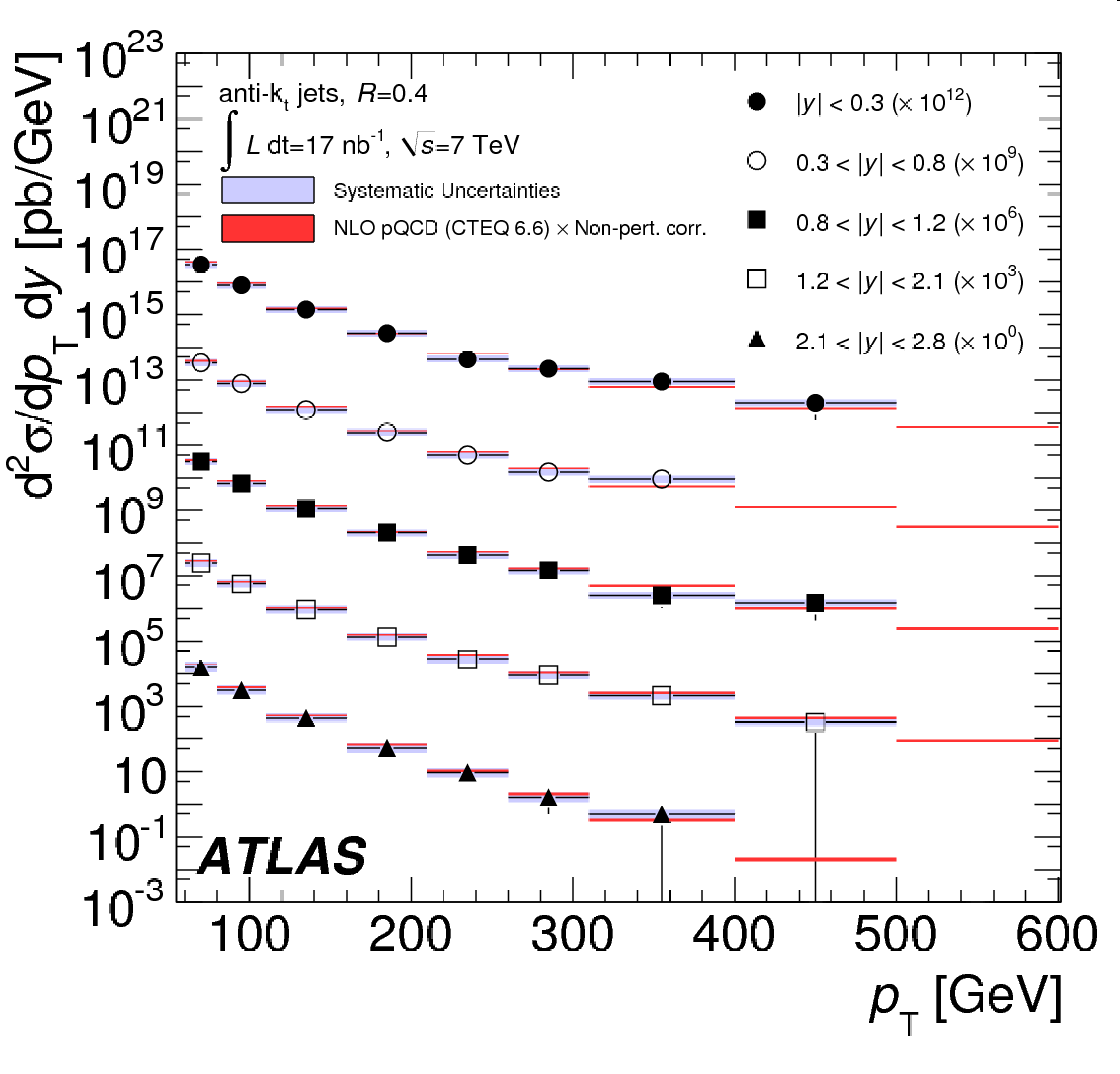}
\end{minipage}
\begin{minipage}{17cm}
\vspace{-7.85cm}
\hspace{2cm}{\Large a)}\hspace{11cm}{\Large b)}
\vspace{5cm}
\end{minipage}
\vspace{-0.7cm}
\caption{a) The particle azimuthal separation with respect to the 
leading particle in MB interactions. b) The 
inclusive single jet cross section, doubly differential in 
the jet $p_T$ and rapidity.}
\label{atlasfigone}
\end{figure}

The charged particle multiplicity data from the analysis of minimum bias (MB)
events~\cite{atlasone,atlasthree,atlasfour} have already, and will 
continue to prove invaluable for the study of soft QCD and the underlying event (UE).
Figure~\ref{atlasfigone}a shows the azimuthal separation between the particles 
and the leading particle for MB events for different requirements on 
the leading particle  $p_T$. The increased collimation of the event
with increasing leading particle $p_T$ illustrating the onset of 
hard QCD can be clearly seen. For the Monte Carlo tune shown, the particle multiplicity 
transverse to the leading particle is too low and the MC events themselves 
appear more collimated than the data. Figure~\ref{atlasfigone}b shows the inclusive single jet
cross section 
doubly differential in 
the jet $p_T$ and rapidity. Within the large uncertainties, the NLO prediction
corrected for hadronisation and UE describes the data well over 
five orders of magnitude.




The LHC has been working well and after less than a year
of collisions at 7~TeV is already providing a large range of 
valuable physics data. Despite the large range of 
high quality results already available from the ATLAS Collaboration, 
these only begin to explore the available phase space. 
Many analyses are still statistically limited in the most interesting
regions of phase space and the Collaboration is working hard to 
reduce the systematic uncertainties. Given the status of the 
statistical and systematic uncertainties, perturbative QCD appears to 
be in reasonable shape.

\tit{First CMS results}

\auth{Thomas Hebbeker\footnote{On behalf of the CMS collaboration.} (RWTH, Aachen)}

In the year 2009 the LHC collider at CERN started with proton proton
collisions at a center of mass energy of 900 GeV, later the energy was
increased to 2.36 TeV and in March 2010 a 7 TeV run began 
which ended in November 2010. The CMS experiment\cite{cms} 
has recorded about 40/pb of 
integrated luminosity at this record energy. The CMS detector performed very
well and many interesting measurements were made. 
Lead-lead collisions at a total center of mass
energy of 574 TeV at the end of 2010 brought new insights into 
heavy ion physics.
First of all the CMS collaboration has `re-discovered' all
Standard Model particles, including W and Z bosons decaying
leptonically, the heaviest quark top and the lighter quarks
in form of various meson and baryon resonances. 
These analyses demonstrate that the 
detector has reached the design values for efficiency and resolution.
The inclusive jet\cite{cms-jets} and dijet\cite{cms-dijets} 
production cross sections were
among the first CMS measurements. The figure\cite{cms-jets}
shows that QCD calculations
can reproduce the jet yields very well. 

\begin{figure}[htpb]
\centering
\includegraphics[height=7.5cm,angle=0]{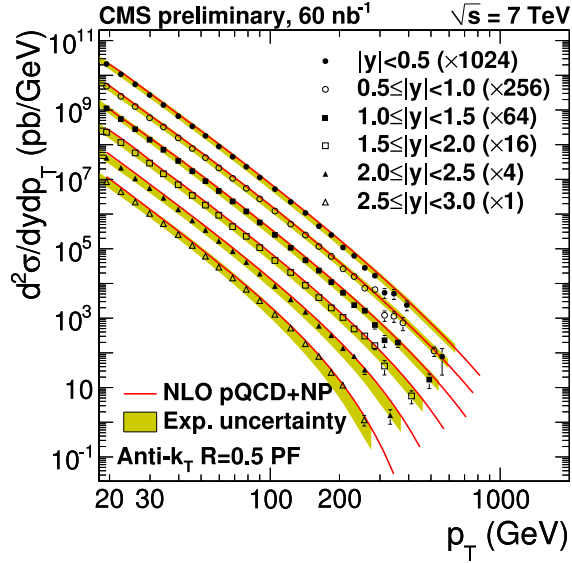} 
 \vskip -0.5cm
\caption{Jet spectra measured at various rapidities in $p p$ at 7~TeV compared to NLO pQCD.}
\end{figure}

The production of charged particles in minimum bias events 
(more precisely: in Non Single Diffractive 
events) was studied
in great detail, in the pseudorapidity range $|\eta| < 2.4$
and for transverse momenta as low as 30~MeV/c.
In particular the yield as a function of $\eta$, the multiplicity
distribution and the $p_T$ distribution were determined 
at 0.9 TeV, 2.36 TeV and 7 TeV and compared
to different model predictions and to data at other center of mass
energies\cite{cms-rapidity}.
Furthermore the production of strange particles like
the $\Xi^-$ were measured\cite{cms-strange}.
Overall the increase of cross sections and multiplicity with center of mass
energy is steeper than anticipated. The current models
with parameters tuned without using LHC data do not provide a 
satisfactory description in all details. 
Also the shape of jets and the topology of hadronic events in general were
analysed in detail\cite{cms-jetshapes,cms-eventshapes}.
The models \pythia\ and \herwig\ provide a good
description of these CMS measurements. 
A very interesting new feature was discovered in $p\,p$ events with a very high
charged particle multiplicity ($N_{_{ch}} > 110$, $p_T = 1-3$ GeV/c). In the 
two-particle correlation as a function of $\Delta \eta$ and $\Delta \phi$
a `ridge' structure, a long range correlation in $\Delta \eta$ 
at small values
of the azimuthal distance $\Delta \phi$ was revealed by CMS\cite{cms-ridge}. 
Current Monte Carlo models cannot explain this feature. 
Finally already after a few days the first interesting heavy ion 
results were made public by CMS\cite{cms-talk}: 
Z boson production in lead-lead collisions
and `jet quenching', seen as dijet events with very different 
energies carried by the two jets.

Beyond QCD results, from the smooth falloff of the measured dijet cross section 
as a function of dijet mass one can set limits on new particles.
For example excited quarks can be excluded within a contact interaction model 
up to a mass of 1.58 TeV at 95\% confidence level\cite{cms-dijets}, 
thus improving 
older Tevatron limits significantly.
From more than 100'000 W decays and about 10'000 Z decays their
production cross sections were measured, these results are 
in good agreement with NNLO QCD calculations\cite{cms-wz}. 
Also a first measurement of the top cross section was made\cite{cms-top}, 
confirming the expected strong rise 
(by a factor of about  25) with respect
to proton - antiproton collisions at 2 TeV center of mass energy.

\tit{Charged-particle multiplicity in $p p$ and heavy-ion collisions at collider energies}

\auth{Jan Fiete Grosse-Oetringhaus (CERN) }

The topical review 
\cite{jfgo_review} summarizes and critically reviews measurements of charged-particle multiplicity
distributions and pseudorapidity densities in pp($\pbar$) collisions between $\cms$ = 23.6~GeV 
and 1.8~TeV. Related theoretical concepts are briefly introduced: Feynman
scaling which is based on phenomenological arguments about the exchange of quantum numbers predicts
that the average number of charged particles increases with $\log \cms$ which implies that
the rapidity density $d\nch/dy$ is a constant as function of $\cms$. Feynman scaling is not fulfilled
in the measured energy range. KNO scaling postulated by Koba, Nielsen and Olesen in 1972 asserts 
that the multiplicity distribution falls onto a universal curve when
rewritten as $P(\nch) \rightarrow 1/\expval{\nch} P(z)$ with $z = \nch / \expval{\nch}$. KNO
scaling is valid for NSD (non single diffractive) collisions in full phase space up to SppS
energies ($\cms$ = 200~GeV). For limited $\eta$-intervals it still remains valid at LHC
energies~\cite{alice_mult1,cms-rapidity}. Negative binomial distributions (NBDs) describe
multiplicity distributions well, a fact which is theoretically not well understood. Cluster models
which assume an independent emission of clusters followed by stimulated particle emission within a cluster
lead to multiplicity distributions of NBD type. NBDs describe
multiplicity distributions of NSD collisions in full phase space up to $\cms$ = 900~GeV
where deviations have been observed. In limited $\eta$-intervals the description still holds at LHC
\cite{alice_mult1,alice_mult2}. The combination of two NBDs, one representing soft and the other
semi hard events (defined as events with and without minijets, respectively) describes multiplicity
distributions up to the highest measured energy in full phase space (1.8~TeV) and in
limited phase space intervals also at the LHC. The identification of trends of the fit parameters
turns out to be ambiguous and assumptions are needed to obtain a coherent picture. 

Although no sound theory arguments exist at present why multiplicities in pp and e$^+$e$^-$
collisions should behave similarly, the average multiplicities as function of $\cms$ are similar when
the concept of effective energy is introduced: $E_{\rm eff} = \cms - E_{\rm lead,1} - E_{\rm
lead,2}$ where $E_{\rm lead,i}$ is the energy which is retained by the proton remnants. The
available energy for particle production is characterized by the inelasticity $K = E_{\rm eff} /
\cms$. This concept allows one to fit the average multiplicity in pp collisions with the following
form: $f_{\rm pp}(\cms) = f_{\rm ee}(K \cms) + n_0$. $n_0$ is the contribution of the leading
protons to the total multiplicity. One obtains a good fit result with $K = 0.35$ and $n_0 = 2.2$, i.e.,
about one third of the energy is available for particle production compared to e$^+$e$^-$
collisions. Phenomenologically, one could ask if this is a hint that only one out of the three
valence quarks is available for particle production. However, the similarity between pp and
e$^+$e$^-$ collisions cannot be found in more differential distributions, e.g. in (pseudo)rapidity
densities. 

Results from LHC show a faster increase of the average multiplicity than anticipated by models
\cite{alice_mult2,cms_mult1}. In particular phenomenological extrapolations and MCs with
pre-LHC tunes underpredict the results at 7~TeV. Updated tunes which include LHC data indicate
that parameters governing the amount of multiple-parton interactions need to be modified to
reproduce the LHC multiplicities. 

\tit{LHCb QCD results}

\auth{Sebastian Schleich\footnote{On behalf of the LHCb collaboration.} (TU Dortmund)}

The Large Hadron Collider (LHC) delivered data in proton-proton collisions at unprecedented 
center of mass energies of $\sqrt{s}$~=~900~GeV and 7~TeV, 
which allow one to test quantum chromodynamics (QCD) predictions in both, the perturbative and the 
non-perturbative regime. For example, the hadronization process falls into the latter. 
Predictions of the hadronization are based on phenomenological models, that are mostly 
tuned on LEP data and their validity under LHC conditions needs to be confirmed by experiments.
Designed for precision measurements in the B meson system, the Large Hadron Collider beauty (LHCb) 
experiment~\cite{LHCBDETECTOR} has several features that account for unique opportunities for QCD 
studies in proton-proton collisions at the LHC: It covers a large forward rapidity range of $1.9<\eta<4.9$. 
Further, its tracking system includes a silicon tracker in close vicinity to the proton-proton interaction 
point and the experiment is equipped with a dedicated particle identification system based on
two ring imaging \v{C}erenkov detectors. 

The $K_S$ production cross section measured at \SSsqrtsL in the kinematic region 
$(p_T <$ 1.6~GeV/c) $\times (2.5 < y < 4.0)$ is found~\cite{Aaij:2010nx} 
to be in agreement with the Monte Carlo prediction (\textsc{Pythia} 6.4, Perugia 0 
tune~\cite{SkandsPerugia}, in the following referred to as MC), the \SSpt spectrum tends 
to be slightly harder on data as compared to MC. Not as good agreement is found in inclusive 
$\phi$ production cross section studies\footnote{All results, except for the $K_S$ cross section 
at \SSsqrtsL are preliminary} at \SSsqrtsH. The measured production cross section in the kinematical 
range (0.8 GeV/c $< p_T <$ 5~GeV/c) $\times (2.44 < y < 4.06)$ 
is significantly enhanced with respect to MC, where also the \SSpt spectrum is harder on data than on MC.\par

The baryon suppression in $\bar\Lambda / K_S$ is a sensitive test of fragmentation models because 
the initial state is purely baryonic. Similar considerations hold for particle-antiparticle ratios, 
since they probe the baryon transport from the beam to the final state. The particle ratio measurements 
$\bar\Lambda / K_S$, $\bar\Lambda / \Lambda$ and $\bar p / p$ are presented at both, \SSsqrtsL and \SSsqrtsHs. 
The $\bar\Lambda / K_S$ ratio is underestimated by MC at both beam energies. In contrast, 
the $\bar\Lambda / \Lambda$ ratio in the kinematical range $2 < y < 4$ is significantly overestimated 
by MC at \SSsqrtsL, whereas at \SSsqrtsHs the data is in better agreement, but still slightly lower than MC.  
The $\bar p / p$ ratio, measured in the range $2.8 < y < 4.5$, is slightly lower on \SSsqrtsL data than on 
MC, whereas it is in rather good agreement at \SSsqrtsHs.
The hard QCD measurements presented are the Drell-Yan muon production \SSpt spectrum at \SSsqrtsH, which is 
found to be in good agreement with the Monte Carlo prediction based on MCFM NLO. Additionally, the lepton 
\SSpt spectrum in in the $W\rightarrow \mu \nu$ decay, as well as the charge asymmetry versus lepton 
pseudorapidity in this decay channel is presented. The results presented, still based on a relatively 
small data sample (59~nb$^{-1}$), are in agreement with Monte Carlo predictions 
within their statistical uncertainties.

\tit{First TOTEM results and perspectives}

\auth{Karsten Eggert\footnote{On behalf of the TOTEM collaboration.} (CERN \& Cleveland) }

TOTEM is a dedicated experiment focused on forward physics  complementary to the programmes of 
the large general-purpose experiments at the LHC~\cite{totem}.  
TOTEM will measure the total proton-proton cross-section with the luminosity independent method based 
on the Optical Theorem which requires a detailed study of the elastic scattering cross-section down to 
a squared four-momentum transfer $|t|$ of 10$^{-3}$ GeV$^2$ and the measurement of the total inelastic rate. 
Furthermore, TOTEM's physics programme aims at a deeper understanding of the proton structure  by 
studying elastic scattering at large momentum transfers, and via a comprehensive menu of diffractive processes.
To perform these measurements, TOTEM requires a good acceptance for particles produced at very small 
angles with respect to the beam. The coverage in the pseudorapidity-range of 3.1 $< |\eta| <$ 6.5 ($\eta = -\ln\tan\theta/2$)
on both sides of the intersection point is accomplished by two telescopes for inelastically produced charged 
particles and complemented by Silicon detectors in special movable beam-pipe insertions -- so called Roman pots
(RP) -- placed at about 147~m and 220~m from the interaction point, designed to detect leading protons at merely 
a few mm from the beam centre.

During the year 2010, TOTEM has participated at the normal low-$\beta^\star$ high-intensity
runs with the vertical RP detectors at a distance to the beam of 18$\sigma$ and the horizontal at 20$\sigma$. 
With this configuration, large-$t$ elastic scattering could be measured for $t$- values above 2.2 GeV$^2$. 
Furthermore, Single Diffraction and Double Pomeron processes were detected over a large region in the 
($\xi = \Delta p/p$, $t$) plane and correlations between the forward proton and the particle densities in 
the very forward inelastic detectors are studied as a function of $\xi$. In addition, data were taken with 
only a few bunches during TOTEM dedicated runs where the vertical Roman Pots have been moved closer to the 
beams (7$\sigma$) and a low intensity bunch with ~10$^{10}$ protons, to reduce pile-up, was added. 
Out of this data sample of several million events, 80k elastic scattering events with $t$-values above 
0.4 GeV$^2$ were extracted. The $t$-distribution showed the usual exponential  slope at low $t$, but also 
exhibits a clear diffractive dip at around 0.6 GeV$^2$, as it was observed for the first time at the ISR, 
whereas proton-antiproton elastic scattering only showed a shoulder in the $t$-distribution. 
The combination of these dedicated runs with the standard high-intensity runs will allow TOTEM to measure 
the $t$-distribution in the range of 0.4 - 4 GeV$^2$ and to distinguish between the various models on the
market. The low intensity bunch with a reduced pile-up of about 1\% is used to 
measure the forward charged multiplicity and forward-backward multiplicity correlations.
During the year 2011, TOTEM will concentrate on the total cross-section measurement, which becomes possible 
with the presently installed inelastic Cathode-Strip-Chambers. However, a special beam optics with large $\beta$* 
around 90 m has to be developed to enable the measurement of sufficiently small $t$-values necessary for the 
extrapolation to the optical point. Furthermore, the extensive studies of the forward particle flow and 
diffractive topologies will continue.

\tit{LHCf results}

\auth{Massimo Bongi\footnote{On behalf of the LHCf collaboration.} ~(Universita \& INFN, Firenze)}

\begin{figure}[htpb]
\centering
\includegraphics[width=0.55\textwidth]{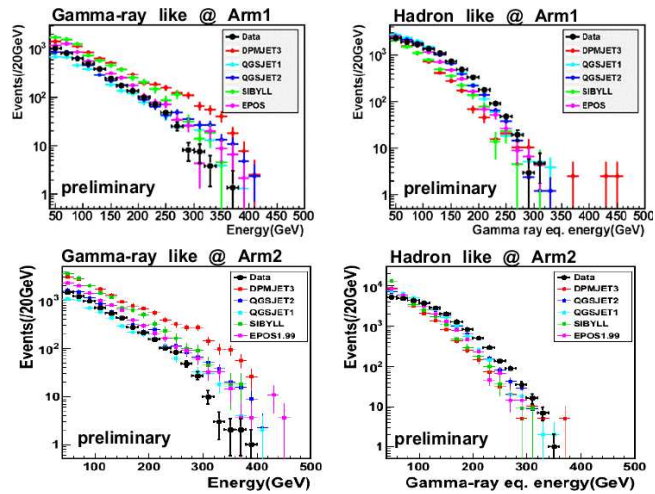}
 \vskip -0.2cm
\caption{Energy spectra of $\gamma$- and hadron-like events measured by LHCf 
Arm1 (top) and Arm2 (bottom) in $pp$ at $\sqrt{s}=0.9$~TeV compared with 
MC simulations.}
\label{fig:lhcf}
\end{figure}

LHCf is an LHC experiment designed to study the very forward production of neutral particles in $p p$ collisions.
Its results can provide valuable information for the calibration of the hadron interaction models used in 
Monte Carlo simulation codes, aiming in particular to clarify the interpretation of the energy spectrum 
and the composition of high energy cosmic rays as measured 
by air-shower experiments. The highest-energy data currently available for the forward neutral-pion 
production spectrum reach $\sqrt{s}=630$~GeV (UA7 experiment~\cite{UA7-1990} operated at 
the Sp$\bar{\mathrm p}$S in 1985-1986).
The LHCf set-up consists of two imaging calorimeters (Arm1 and Arm2) symmetrically placed 140~m away on 
both the sides of the ATLAS interaction point, covering the pseudorapidity range $|\eta|>8.4$.
Further information about the scientific goal, the technical details and the performance of the detectors 
can be found in the following references:~\cite{TDR-2006,NIM578-2007,ACTA38-2007,JINST3-2008,NIM612-2010,JINST5-2010,APH-2010}.
The experiment has successfully finished taking $p p$ data at $\sqrt{s}=0.9$~TeV and at $\sqrt{s}=7$~TeV in 2009-2010.
A preliminary analysis of the energy spectra of $\gamma$-ray like and hadron like events measured by the 
calorimeters (Arm1 - upper plots, Arm2 - lower plots) at $\sqrt{s}=0.9$~TeV, compared with the expectation 
of MC simulations, are shown in the Fig.~\ref{fig:lhcf}. Only statistical errors are reported and simulations are
normalized  by total entries of $\gamma$- and hadron-like events. The detectors were removed 
in July 2010 for an upgrade which will improve their radiation hardness, and they will be 
back in the LHC tunnel for the collisions at $\sqrt{s}=14$~TeV.

\tit{Tevatron results of relevance for cosmic-rays}

\auth{Lars Sonnenschein\footnote{On behalf of the CDF and D\O\ collaborations.} (RWTH Aachen) }

The two multi-purpose experiments D\O~\cite{d0_det} and CDF~\cite{cdf_det} are operated 
at the Tevatron collider,
where proton anti-proton collisions take place at a centre of mass energy of 1.96~TeV in Run~II.
In the kinematic plane of $Q^2$-scale and (anti-)proton momentum fraction $x$, Tevatron jet
measurements cover a wide range, with phase space regions in common and beyond the HERA 
$ep$-collider reach. 
The kinematic limit of the Auger experiment is given by a centre of mass energy of about
100~TeV. Cosmic rays cover a large region of the kinematic
phase space at low momenta 
$x$, corresponding to forward proton/diffractive physics and also at low scales,
corresponding to the hadronisation scale and the underlying event.
Therefore of particular interest are exclusive and diffractive measurements
as well as underlying event, double parton scattering and minimum bias
measurements.
The kinematic limit of the Tevatron corresponds to the PeV energy region below the knee
of the differential cosmic particle flux energy distribution.
The data discussed here are in general corrected for detector effects, such as efficiency and acceptance.
Therefore they can be used directly for testing and improving existing event generators and 
any future calculations/models. Comparisons take place at the hadronic final state
(particle level).
 
In particular for elastic and exclusive production 
measurements~\cite{d0_6065_2010,cdf_prl102_2009} 
forward proton detectors,
which cover a pseudorapidity range of up to $|\eta|\lesssim 8$ and momentum fractions
of $0.03<\xi\lesssim 0.10$
are useful, to detect the intact (anti-)proton. 
Many further analyses of exclusive and diffractive
production~\cite{cdf_arXiv1007.5058_2010,d0_6042_2010,cdf_prd77_2008,cdf_prelim_2006,cdf_prl102a_2009,cdf_prl_99_2007,cdf_prl_98_2007}
have been accomplished.
Underlying event, double parton scattering and minimum bias studies have been addressed
by the measurements
\cite{d0_6054_2010,cdf_10084_2010,d0_prd81_2010,cdf_prd82_2010,cdf_prelim_2010,cdf_prl102b_2009}.
The studies have been pioneering work in many cases.
Methods have been established which are widely used by LHC experiments today.
The measurements have provided very important input to theorists,
in particular with respect to non-perturbative QCD physics, where phenomenological models
are varying considerably.
Most prominently the breakdown of factorisation between HERA and Tevatron
has been established~\cite{cdf_prd77_2008}.
The double Pomeron exchange mechanism offers the possibility to study the exclusive Higgs production
at the LHC, where predictions did vary by a factor of 1000 before the CDF 
measurement~\cite{cdf_prd77_2008}. Already in Run~I CDF has provided useful input for diffractive 
parton distribution functions.

\tit{QCD results from HERA}

\auth{Armen Bunyatyan\footnote{On behalf of the H1 and ZEUS collaborations.} (Yerevan \& DESY)}

New QCD results obtained by the H1 and ZEUS experiments at HERA collider are reviewed.
These results are based on data taken in $e^{\pm}p$ collisions  during 1994-2007
corresponding to an integrated luminosity of almost
$\rm 0.5~fb^{-1}$ for each experiment.
HERA provides unique information on the proton structure. High center of mass
energy $\sqrt{s}=320$~GeV gives access to both the low Bjorken-$x$ domain
and regime of high momentum transfers squared $Q^2$.
An ultimate precision of DIS cross section measurement is achieved by
combining the H1 and ZEUS measurements.
The combined data are used as a sole input to a QCD fit to obtain HERAPDF sets.
Important cross checks of the conventional QCD picture and the additional constraints
for the gluon density distribution in the proton  are provided by the
measurements of the structure function $F_L$ and the charm and beauty
cross sections.
New measurements of inclusive neutral and charged current scattering 
cross sections at high $Q^2$ improve precision in this kinematic domain.
Jet production at HERA provides an important testing
ground for pQCD and new constraints for the gluon density distribution
in the proton.
The running of strong coupling $\alpha_S$ is demonstrated and its
value at $Z^0$ mass, $\alpha_S(M_Z^0)$, is determined from the jet
measurements at HERA with high precision.

The cross section of inclusive DIS diffractive process is measured
within a wide kinematic range.
The diffractive parton distribution functions (DPDFs) of the proton
are determined from  QCD fits to the data
 including the dijet production cross sections in diffractive DIS.
Predictions based on these DPDFs are in agreement with the
measured cross section of diffractive dijet and charm production in DIS
at HERA and the longitudinal diffractive structure function $F^D_L$.

Data from the recent measurements of leading proton and neutron production
are presented and compared to the theoretical models.
The measurements are well described by the models
which include the  baryon production via virtual meson exchange.
The leading proton and neutron data from the H1 and ZEUS experiments
are also compared with the hadron interaction models which are used 
in the analyses of ultra-high energy cosmic rays.
The sensitivity of the HERA leading baryon data to the differences between
the models is demonstrated.


\tit{Performance of the ALICE experiment for cosmic ray physics}

\auth{Bruno Alessandro\footnote{On behalf of the ALICE collaboration.} (INFN, Torino)}

A large number of atmospheric muon events were recorded during 2009
and 2010 for the calibration, alignment and commissioning 
of most of the ALICE (A Large Ion Collider Experiment at the CERN LHC) 
detectors. In this Workshop we presented the analysis of some
of these data to understand the performances and the possibilities of 
ALICE to study topics connected to cosmic ray physics.
The ALICE central detectors select atmospheric muons with zenith angle 
in the range $0^{\circ}-60^{\circ}$. The muons are tracked in a large volume Time 
Projection Chamber (TPC) that measures the muon multiplicity, 
and for each muon the momentum, the sign, the direction and 
the spatial coordinates. An analysis  of these observables and 
some correlation among them have been presented. In particular 
the muon multiplicity distribution and some events with very 
high multiplicity have been shown and detailed analyzed.
A first attempt to measure the ratio $\mu^+/\mu^-$ for vertical muons
($0^{\circ}-20^{\circ}$) with a limited statistics has been presented and compared
with world previous measurements. 

Horizontal  muons, that is muons with zenith angle  
in the range $60^{\circ}-90^{\circ}$ are very rare events that have been detected 
by the Forward Muon Spectrometer in 9 days of data taking. 
A selection of these events to obtain a good sample to measure
the momentum distribution and the ratio $\mu^+/\mu^-$ at surface
level has been discussed and measurements shown.

\tit{CMS results on forward physics and other of relevance for cosmic rays}

\auth{Lev Kheyn\footnote{On behalf of the CMS collaboration.} (MSU, Moscow)}

Evidences of observation of single-diffraction at the LHC are presented at 900 and at 2360 GeV.
Single-diffractive events appear as a peak at small values of the variable $E \pm pz$ which is
proportional to $\xi$, the proton fractional energy loss, reflecting the $1/\xi$ behaviour of
the diffractive cross section. Single-diffractive events also appear as a peak in the
energy distribution  of the forward calorimeter HF, reflecting the presence of a rapidity
gap over HF. The data have been compared on the detector level to \pythia\ 6 and \phojet\
generators. \pythia\ 6 gives  a better description of the non-diffractive component of the
data, while \phojet\ reproduces the  diffractive contribution more accurately.

The energy flow (at detector level) for minimum bias events and events having a hard scale
defined by a dijet with $E_{T,jet} >$8 GeV ($E_{T,jet} >$20 GeV for $\sqrt{s}$ = 7~TeV)
in  $|\eta| <$ 2.5 has been measured for the first time in hadron-hadron collisions in the
forward region of 3.15 $< |\eta| <$ 4.9. The increase in energy flow in the forward region
with increasing centre-of-mass energy is significant and is reproduced by Monte Carlo (MC)
simulations for events with dijets, whereas it is underpredicted for minimum bias events.
None of the MC simulations manages to describe all energy flow data. Monte Carlo
tunes which are closer to the measurements for minimum bias events differ from measurement
in dijet events. Particularly for minimum bias events, the measured energy flow at
$\sqrt{s}$ = 7~TeV is larger than any of the predictions. Those MC simulations
which best describe the energy flow in the forward region are
different from those which best describe the complementary measurements of charged
particle spectra in the central region. The measurement of the energy flow in the forward
region therefore provides further input to the tuning of MC event generators and
it constrains modelling of multiple interactions at high energies. 
CMS measurements of charged particle pseudorapidity densities are being compared with 
cosmic ray (CR) generators. The distribution proves of high discriminative power,
revealing large spread of the CR MCs predictions at highest energy of 7~TeV.

The flux ratio of positive- to negative-charge cosmic muons has been measured as a function
of the muon momentum and its vertical component. This is most precise measurement below 100
GeV. The ratio was measured over broad range 10 GeV - 1 TeV of transition from
approximately constant to a rising value.

\tit{STAR results of relevance for cosmic rays}

\auth{Joanna Kiryluk\footnote{On behalf of the STAR collaboration.} (LBL, Berkeley)}

The Relativistic Heavy Ion Collider (RHIC) is a versatile accelerator situated at 
Brookhaven National Laboratory which commenced operations in 2000. 
It collides heavy ions as well as polarized protons at center of mass energies of up to 200\,GeV 
per nucleon and 500\,GeV, respectively.
The Solenoid Tracker At RHIC (STAR) detector\,\cite{star-detector} provides tracking, particle 
identification, and electromagnetic calorimetry covering large acceptance.
Key strengths include the capability to reconstruct jets, study correlations, and identify 
particles in high multiplicity environments. 

Measurements in unpolarized $p p$ collisions at RHIC test perturbative QCD (pQCD) calculations. 
The data on jet and inclusive particle production cross sections~\cite{pp,dAu-star} are in good 
agreements with Next-to-Leading-Order (NLO) pQCD calculations. 
Recent measurements of the $W^{\pm}$ boson production in 500\,GeV $p p$ collisions at RHIC and in 
7\,TeV collisions at the LHC are in good agreement with NLO pQCD 
over a wide range in $\sqrt{s}$\,\cite{pp-w}. 

The flux of prompt leptons at the Earth is of importance to cosmic ray (CR) and neutrino physics.  
Estimates depend strongly on models for the charm cross section and energy spectra.
These models use the pQCD framework and extrapolate charm collider data to CR energies.
The flux of prompt leptons is strongly dependent also on the small-$x$ nuclear gluon distributions.
New dynamical effects, such as parton saturation that may be observed at forward collider 
rapidities, change the  DGLAP dynamics and thus the flux estimates\,\cite{prompt}. 
STAR and PHENIX data~\cite{pp-non-photonic} on electrons from heavy-flavor decays
are consistent in the regions of kinematic overlap and are well described by Fixed-Order-Next-to-Leading 
Logarithm calculations~\cite{fonll}.
The STAR J/$\Psi$ data\,\cite{pp-J-Psi}  are well-reproduced by calculations\,\cite{NRQCD} using the 
color octet and singlet models in  non-relativistic QCD.
STAR has determined the B-hadron feed-down contribution to the inclusive $J/\Psi$ yield from 
$J/\Psi$-hadron azimuthal angle correlations.
It is found to be 10-25\% and has no significant $\sqrt{s}$ dependence from RHIC to LHC 
energies\,\cite{J-Psi-correlation}. The STAR $\Upsilon$(1S+2S+3S) production cross section~\cite{pp-Upsilon} 
is consistent with world data and NLO pQCD calculations in the Color Evaporation Model\,\cite
{CEM}.

Collisions of $d Au$ ions at $\sqrtsnn=200$~GeV at RHIC have made it possible to study the modification 
of elementary QCD processes in cold nuclear matter (CNM), 
and provide insight in coherence effects or shadowing in nuclei, the saturation of small-$x$ gluons, 
parton energy loss, and soft multiple scattering effects.
Forward particle production is found to be suppressed\,\cite{dAu-star,dAu-brahms} and back-to-back 
correlations are reduced\,\cite{di-hadron}, consistent with saturation models\,\cite{CGCsat}.
Hot matter effects have been studied at RHIC in $Au Au$ collisions.
The observation of phenomena such as jet quenching and collective motion 
suggests that a thus far unobserved state of hot and dense matter with partonic degrees of freedom, 
resembling an ideal and strongly-coupled fluid, has been created\,\cite{QGP}.
The ongoing Beam-Energy Scan program aims to observe the anticipated critical point in the QCD phase diagram\,\cite{BES}.
STAR has also observed anti-hypertriton production in $Au Au$ collisions\,\cite{anti-matter}, the first 
ever observation of an anti-hypernucleus. The production and properties of antinuclei, and nuclei 
containing strange quarks, have implications spanning nuclear and particle physics, astrophysics, and cosmology.

\newpage

\section{Hadronic collisions at multi-TeV energies: Theory}

\tit{Monte Carlo tuning at the LHC}

\auth{Holger Schulz\footnote{On behalf of the Professor and Rivet collaborations.} (Inst. f\"ur Phys., Humboldt-Univ. Berlin)}

Monte Carlo simulations of high energy physics processes are essential for
many aspects of the LHC physics programme, e.g. the experiments use them to
determine backgrounds to signal processes and to estimate reconstruction
efficiencies, which are sources of systematic uncertainties that clearly
dominate over statistical errors at the LHC.
In order to reduce these systematic uncertainties, proton-proton collisions
need to be simulated in such a way that they look as close to real data as
possible. This however can in many cases only be achieved by optimising
phenomenological model parameters to data, especially when 'soft' QCD effects
such as multiple parton interactions or hadronisation are to be described.

An overview of such parameter tuning strategies currently applied at the LHC
has been presented, i.e. conventional manual parameter optimisation by the CMS
collaboration and systematic tunings within ATLAS. The focus was clearly on the
latter, where the software packages Professor and Rivet that allow for a
systematic tuning effort using statistical techniques are used extensively. 

Rivet is an application that reads in generator independent events (``HepMC''
format), processes these events by applying user written routines that mimic
actual data analyses. After processing the generated events, histograms are
produced that use the same binning as published data.

Running a Monte Carlo event generator can be regarded as calculating a very
expensive function. The key feature of Professor is the parameterisation of
this expensive function by means of parameterisations using polynomials such
that one effectively produces a fast analytic model of the Monte Carlo
generator response to shifts in parameter space. It is thus possible to get a
very good approximate description of a generator at a certain point in
parameter space in less than a second; a task that takes hours or even days
with conventional methods.
With this fast model, the task of tuning parameters is therefore passed on to
constructing a goodness-of-fit measure between the parameterised generator
response and real data. In this context, the necessity to have data, corrected
for detector effects has been stressed. 

Further uses of the parameterisation like calculating the sensitivity of
observables to shifts in parameter space and an interactive Monte Carlo
simulator (``prof-I'') have been presented.
The successful usage of Professor and Rivet has been illustrated by recent
tunings performed within the ATLAS collaboration. Examples have been given
(``AMBT1'', ``AUET1'') and it has been stressed how fast the turn-around from
taking new data to getting new tunings that include these data can be. Also,
more special uses of Professor to study systematic variations
(``Eigentunes'', re-tuning using different PDFs) have been presented.

\tit{Parton correlations and fluctuations}

\auth{G\"osta Gustafson\footnote{In collaboration with C. Flensburg, L. L\"{o}nnblad, and A. Ster.} (Lund University)}

Multiple interactions and diffraction are important components in high
energy collisions~\cite{GG1,GG2}.
These effects are influenced by correlations and fluctuations in the
parton evolution, and the understanding of these features is therefore
essential for a proper interpretation of data from LHC and cosmic ray
experiments. At high energies parton distribution functions at very small
$x$-values are governed by BFKL dynamics and saturation effects are important.
Mueller's dipole model is a formulation of LL BFKL evolution in transverse
coordinate space. The Lund Dipole Cascade model is a
generalization of Mueller's model, which also includes non-leading
effects from \emph{e.g.} energy-momentum conservation and running
coupling, saturation effects in the cascade evolution, and confinement.
The model is implemented in a MC called {\sc Dipsy}, and in this
talk I use it to study effects of correlations and fluctuations on
double parton interactions and diffraction.

In parton evolution \`{a} la BFKL the gluons are strongly correlated. An analysis 
of double parton distributions shows increased correlations for small $x$ and 
large $Q^2$. A spike develops for small separations between the partons 
in transverse coordinate space. The correlation can also be expressed in terms
of an "effective cross section", which becomes reduced at high 
energies and large $p_\perp$.

In the Good--Walker formalism diffractive excitation is determined by the fluctuations
in the scattering amplitude between different components in the projectile
wavefunction.  In BFKL the proton substructure in terms of a parton cascade 
has large fluctuations and can fill a large rapidity range. An analysis of these fluctuations
reproduces low and high mass diffractive excitation in DIS and $pp$ collisions. For $pp$ scattering the 
fluctuations are suppressed by unitarity constraints, which leads to a breaking of factorization 
between DIS and $pp$.

The model can also be applied to nuclear collisions, and finally  I present some
preliminary results for exclusive final states in $pp$, $pA$, and $AA$ collisions.

\tit{A new model for minimum bias and the underlying event in \sherpa}

\auth{Korinna Zapp\footnote{In collaboration with H. Hoeth, V. Khoze, F. Krauss, A. Martin and M. Ryskin} (Durham)}

Minimum bias events reveal not only the most complete view on the physics at
hadron colliders, but also have an intimate connection to the underlying event
and are thus highly relevant to many high-$p_\perp$ processes. Higgs searches at
the
LHC, for instance, rely largely on event topologies with rapidity gaps. The
feasibility of such measurements depends strongly on the survival probability of
rapidity gaps. Apart from the connection to the underlying event, diffraction as
an important part of minimum bias events is interesting in its own right.

Unfortunately, no model describing soft, semi-hard, diffractive and hard
QCD events has been implemented in a multi-purpose event generator so far.
The Khoze-Martin-Ryskin model\cite{Ryskin:2009tj} is a multi-channel eikonal
model that by summing all multi-pomeron diagrams is capable of describing
elastic and inelastic scattering, low mass and high mass diffractive
dissociation and central exclusive production.

The Monte Carlo realisation relies on the partonic interpretation of the model.
The simulation of elastic scattering is straight-forward, while the inelastic
collisions are more involved. First, the number of exchanged ladders and the
impact parameters of the ladders have to be generated. The emissions from the
ladder are generated using a Sudakov form-factor, that accounts for absorptive
corrections and Regge dynamics.  The colour charge of
the $t$-channel propagators has to be fixed, the singlet exchanges naturally give
rise to rapidity gaps. Finally, the hardest emissions are corrected to pQCD
matrix elements to reproduce the correct high energy behaviour.
The model will be formulated also as a model for the underlying event and become
available as part of the \sherpa\ 1.3 release.

\clearpage

\tit{Multiple partonic interactions with \herwig++}

\auth{Stefan Gieseke (ITP, Karlsruhe)}

The focus of this talk is on the development of a multiple partonic
interaction (MPI) model for minimum bias interactions and the underlying
event in \herwig++.
We briefly summarize the general purpose Monte Carlo event generator
\herwig++~\cite{Bahr:2008pv,Bahr:2007ni,Bahr:2008tx,Bahr:2008tf} before
describing the development of the MPI model
\cite{Sjostrand:1987su,Butterworth:1996zw,Borozan:2002fk} in detail. We
explain the relevance of the main parameters of our model, the inverse
radius $\mu^2$ and $p_T^{\rm min}$ in detail.  The former characterizing
the width of the spatial transverse parton distribution inside a
hadronic projectile and the latter the transverse momentum down to which
a partonic interaction will still be described by perturbative QCD.  The
development of the model was done in several steps which are briefly
explained.
\begin{enumerate}
\item A semi hard model for MPI, containing only interactions above
$p_T^{\rm min}$~\cite{Bahr:2008dy,Bahr:2008wt}.
\item A soft model, also for interactions below $p_T^{\rm min}$~\cite{Bahr:2008wk,Bartalini:2010su,Bahr:2009ek}.
\item An extension of the model to allow for colour reconnections~\cite{christian}.
\end{enumerate}

The final model is shown to describe data from CDF
\cite{Affolder:2001xt,Acosta:2004wqa} and recent non--diffractive
minimum bias and underlying event data from ATLAS at 900\,GeV and 7\,TeV
\cite{atlasone,atlasfour,nchgeq6}.  The residual
energy dependence of the model parameters is briefly discussed and an
outlook to further work on this dependence is given.

\tit{PYTHIA 8 status}

\auth{Torbj\"orn Sj\"ostrand (Lund University)}

The \textsc{Pythia~8} event generator~\cite{TS1} is the C++ successor 
to the Fortran-based \textsc{Pythia~6}~\cite{TS2}, frequently used in 
the study of $pp/p\bar p$ and $e^+e^-$ physics.
One of the main developments is that \textsc{Pythia~8} now 
contains a complete interleaving of multiparton interactions,
initial-state radiation and final-state radiation, in one 
common sequence of decreasing $p_{\perp}$ scales. That is, 
the features at larger $p_{\perp}$ values set the stage for 
the subsequent dressing-up by softer emissions.
The approximate matching of showers to hard-scattering matrix 
elements has been improved for a large set of hard processes~\cite{TS3}.
For showers in QCD processes the first emission is compared
with $2 \to 3$ matrix elements to confirm a reasonable rate~\cite{TS4}.

The traditional multiparton interactions framework is largely 
retained, but some new possibilities are added. One is that it is 
now possible to preselect two separate hard processes in the
same event, to help simulate signals for double parton scattering.
Another is that rescattering, where one parton scatters twice 
(or more) against partons from the other hadron, can now be 
simulated~\cite{TS5}. Unfortunately it is not simple to find a good 
experimental signal for such events.
Other developments include an improved framework for the structure
of diffractive events, a richer mix  of underlying-event processes,
and an updated set of parton distributions. 

Early attempts to tune \textsc{Pythia~8} to minimum-bias data 
gave too much underlying-event activity. The problem has been 
traced to a double counting between some initial- and final-state
radiation. This has now been fixed, and tunes to Tevatron data
have been produced~\cite{TS4}. Unfortunately these tunes underestimate 
the activity observed in some of the early LHC data sets, which 
either may be owing to problems with the generator not reproducing
different cut conditions, or point to some tension in the data.
For now a slightly separate LHC tune has been made. 

\textsc{Pythia~8} does not reproduce the CMS ridge effect,
which thus shows that some physics mechanisms are still missing.
Similarly there are problems e.g. with the particle composition
and Bose-Einstein effects observed at LHC, that hints towards 
collective effects.\\

\tit{Total hadronic cross sections at high energy}

\auth{Rohini Godbole\footnote{In collaboration with A. Grau, G. Pancheri and Y. Srivastava.} (CTS-Bangalore \& CERN)) }

Energy dependence of total hadronic cross-sections is an important subject,
both from a theoretical point of view due to its intimate relation to the
non perturbative  QCD dynamics and also from a phenomenological point of 
view, in the context of making accurate predictions for high energy cosmic 
ray interactions based on the currently available information. In this talk 
I present a summary of the current state of data and model predictions for the
same,  paying particular attention to the Eikonalised Minijet Model (EMM)
supplemented with soft gluon resummation~\cite{Godbole:2004kx}, which tames the 
unacceptably strong energy rise of the EMM.

The $pp, p \bar p$ and $\gamma p$ ($\gamma \gamma$) cross sections, scaled
by a VMD inspired  factor of  $330$ ($(330)^2$), all in fact show an almost 
universal  behaviour, perhaps with a slightly faster rise for the 
$\gamma \gamma$
induced processes~\cite{Godbole:2008ex}. The models have to provide an
explanation of the initial fall, normalisation at (and the position of) 
the minimum and
the subsequent rise. The important issue to be addressed is the dynamics 
responsible for this rise, consistent with the Froissart bound. Then one can 
investigate the impact of these model predictions, extrapolated to 
cosmic ray energies.
There exist different set of {\it fits} to the current data on total 
cross-sections, some of them with a form chosen so that various constraints 
from unitarity and analyticity are automatically satisfied. These have been 
then normally used to obtain the predicted total cross-sections at high 
energies.

In the EMM models the rise of cross-section with energy is driven by the 
minijet cross-section calculated in perturbative QCD (pQCD), its rise with 
energy given by
$\sigma_{minijet} \propto {\frac{1}{p^2_{tmin}}} {\left[\frac{s}{4 p^2_{tmin}}\right]}^{\epsilon}$ where $\epsilon = J-1$,  $J$ being the degree of 
singularity of the gluon density in the proton. This has to be embedded  
in an eikonal formulation, which guarantees unitarity. The eikonalisation 
involves transverse parton overlap function in the two hadrons $A \& B$, 
$
A_{AB} (\beta) = \int d^2b_1 \rho_A (\vec {b_1}) 
\rho_B  (\vec \beta - \vec{b_1})$. This, along with the
minijet (and some parametrisation of soft) cross-section, 
then is used in building the total cross-section, 
$\sigma^{\rm tot}_{pp(\bar p)}=2\int d^2{\vec b} [1-e^{-n(b,s)/2}]$,
with $n(b,s) = A_{AB} (b,s) \sigma (s)$.

Different EMM models differ in the way $A_{AB} (b,s)$ is modeled. In the 
GGPS model~\cite{Godbole:2004kx}, the energy dependent transverse space 
matter distribution is calculated as the Fourier Transform of the transverse
momentum distribution of the partons, which is built through resummation
of soft gluon emissions from the valence quark in the proton (to the
leading order in $\alpha_s(Q^2)$)~\cite{corsetti}.
In this BN EMM formulation, one of the important factor
affecting the $\sigma^{\rm tot}$ is 
behaviour of $\alpha_s(Q^2)$ in the far infrared which is modeled by a form
such that,
$\alpha_s(k_T^2) \rightarrow (1/k_T^2)^p$ as $k_T^2 \rightarrow 0$. 
The requirement that our form of $\alpha_s$ be consistent with a confining 
potential and the singularity is integrable gives $1/2 < p < 1$.  Further, 
in the high energy limit, one can 
show $\sigma^{\rm tot} \simeq (\epsilon \ln s)^{1/p}$ and thus a high energy 
behaviour consistent with Froissart bound attains in our model naturally.
Prediction of $\sigma_{\rm tot}$ over the whole energy range requires also 
the soft 
cross-section, $\sigma^{\rm soft} (s) $ and the corresponding overlap function
$A_{AB}^{BN, soft}$, which are parameterised. 

\begin{figure}[htb]
\begin{tabular}{cc}
\includegraphics*[width=7cm,height=4.5cm]{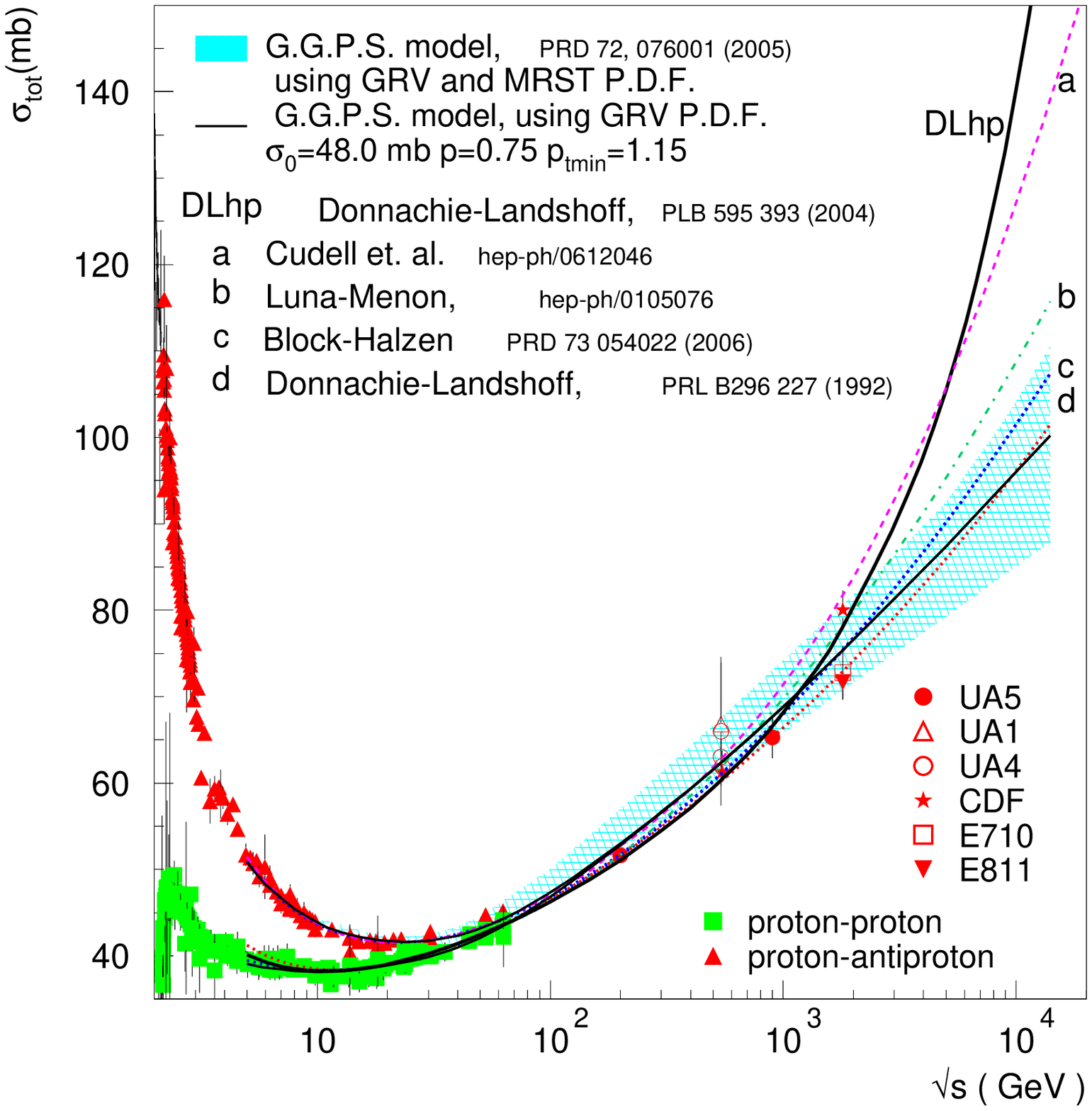}
\includegraphics*[width=7cm,height=4.5cm]{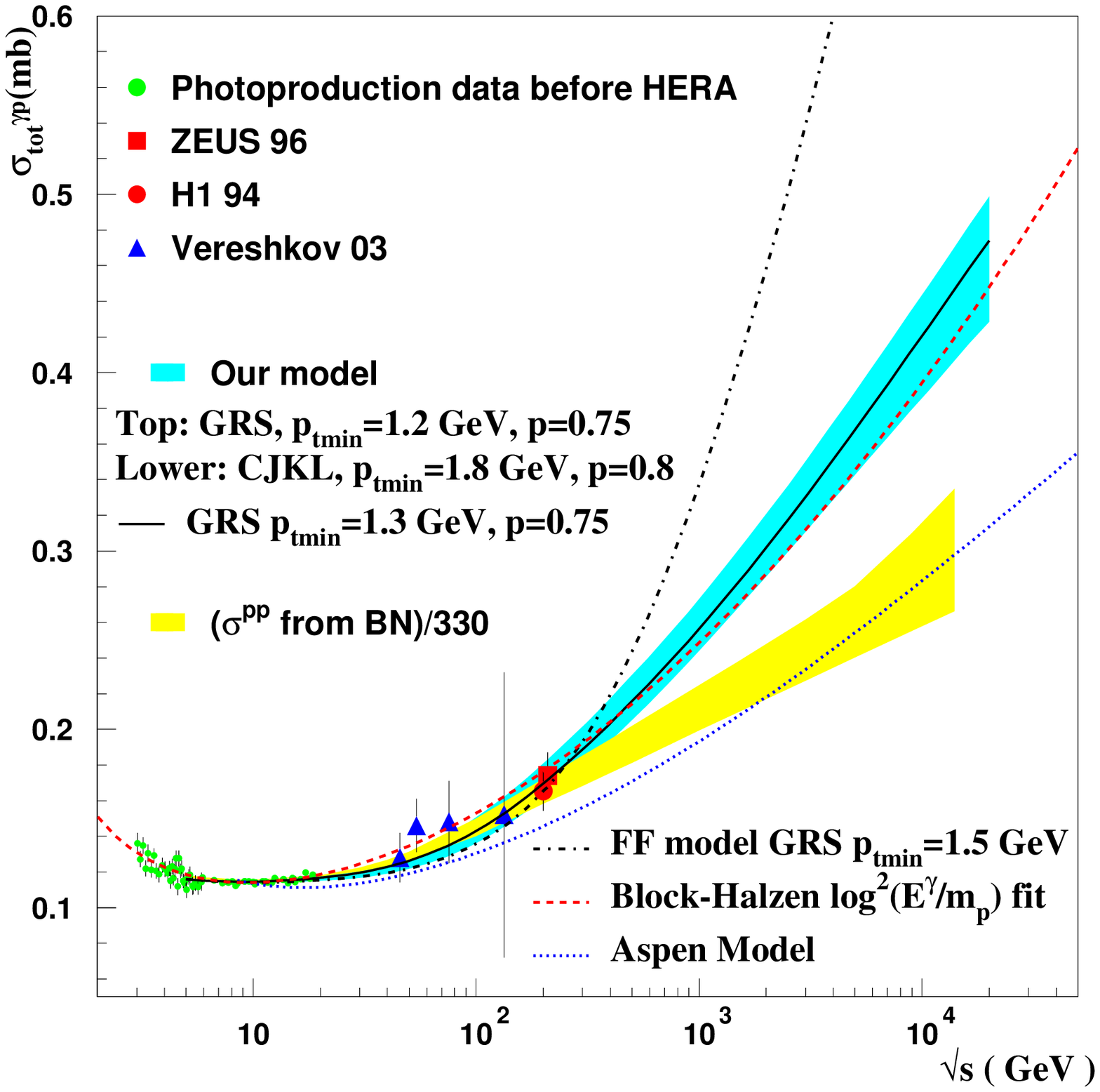}
\end{tabular}
\caption{Predictions of the BN EMM~\cite{Godbole:2004kx} for $pp/p\bar p$ in the
left panel and $\gamma p$ case ~\cite{Godbole:2008ex} in the right panel, 
compared with different other model predictions as well as the data.} 
\label{fig1}
\end{figure}

The left panel of Fig.~\ref{fig1} taken from Ref.~\cite{Godbole:2004kx},
shows comparisons of a variety of model predictions with data and each other, 
the blue band corresponding to the spread in our fits. We see that the 
LHC data will already offer us a nontrivial discrimination among the models. 
For the photon induced processes, the BN EMM has to be supplemented by one
more parameter $P_{\rm had}$, the probability that the photon develops a 
hadronic structure. We find that the best fit seems to prefer a rise of
$\gamma p$ cross-sections a little faster than the soft-pomeron 
predictions as well as those obtained, using factorisation, from our blue band 
for $\sigma_{\rm tot}^{pp}$ of the left panel. 
Of course now the interesting thing is to try and see the effect of these 
range of predictions of $\sigma_{\rm tot}^{pp/\bar p p}$ and the 
somewhat faster rise of the $\sigma_{\rm tot}^{\gamma p}$ on the
cosmic ray simulations. Preliminary investigations~\cite{cornet}, for a photon
energy of $10^{19}$ eV, show that this difference can affect the development
of the longitudinal profile of muonic showers  by $\sim 10 \%$.

The model has been recently applied to $\pi p$ and $\pi \pi$ cross-sections
~\cite{lianew}. In these cases the energy ranges over which the data are
available are rather limited and in the latter case even the onset of the rise 
is not clearly established. 
The LHC with zero degree calorimeter (ZDC) can in fact provide nontrivial 
information to increase our confidence in making predictions for collisions 
with cosmic ray energies.
In conclusion, the different models for total cross-sections seem to be 
good shape. The QCD based BN-EMM, which uses the experimentally measured
parton densities and ideas of soft gluon resummation,  in fact is able to 
even predict the Froissart bound. The same model gives a consistent description
of high energy $\gamma p$ data and can be extended to $\pi$ induced processes.
The LHC data can play an important role in sharpening up our predictions so
that they can be extended to the highest cosmic-ray energies.

\tit{On strategies for determination and characterization of the underlying event}

\auth{Sebastian Sapeta (LPTHE, Paris)}

The underlying event~(UE) is a soft activity which accompanies each hard
process studied at hadron hadron collider. 
Its good understanding is of great importance since the UE affects a
wide variety of high-$p_T$ measurements, e.g. by introducing a bias or by
degrading kinematic jet reconstruction.
However, both an unambiguous definition of UE and its modeling faces a number
of problems. 
Therefore, it is particularly important to be able to measure the UE as well as
possible so that corresponding experimental results can be used as an input to
further constrain the models.
 
We have carried out a twofold study devoted to this
issue~\cite{Cacciari:2009dp}.
First, we asked the question of how the existing methods of UE determination
perform on the practical aspects of the problem.  Developing a simple toy model
of UE and using is as a testing ground, we have examined two methods: the
``traditional approach''~\cite{KarFieldDY} and the more recent area/median 
approach~\cite{Cacciari:2007fd, Cacciari:2008gn}.
One conclusion from this part of our study is that for determinations of
averaged quantities, like the average transverse momentum of UE per unit area in
the (y,$\phi$)-plane, $\mean{\rho}$, both methods give comparably good results.
In contrast, for event-by-event measurements and determinations of fluctuations
of UE, the traditional approach is affected significantly more
by the contamination coming from the hard part of the event.

The second question we studied was that which observables related to the energy
flow of UE are interesting to measure.
Here, we chose the area/median method to examine more realistic UE from the
Monte Carlo (MC) models and found noticeable differences between predictions of
different generators/tunes extrapolated to LHC energy. 
Therefore, we conclude that a broader range of observables deserves dedicated
measurements. Those include rapidity dependence of $\rho$, intra- and
inter-event fluctuations and correlations. A first step of this program has
already bin made by the CMS collaboration which used the area/median approach
to measure the charged component of the UE at $\sqrt{s} = 0.9$ TeV. The
preliminary results~\cite{CMS-UE} show that, even at low multiplicities, the
method is capable to constraint MC tunes.

\tit{Gluon saturation at the LHC from Color-Glass-Condensate}

\auth{Amir Rezaeian (Valparaiso) }

At high energy, a system of parton (gluons) forms a new state of
matter: Color-Glass-Condensate (CGC)~\cite{CGC}. The CGC is the
universal limit for the components of a hadron wavefunction which is
highly coherent and extremely high-energy density ensemble of
gluons. In the CGC picture, the density of partons $\rho_p$ with a
typical transverse momenta less than $Q_s$ reaches a high value,
$\rho_p \propto 1/\as \,\,\gg\,\,1$ ($\as$ is the strong coupling
constant). The saturation scale $Q_s$ is a new momentum scale that
increases with energy. At high energies/small Bjorken-$x$, $ Q_s
\,\,\gg\,\,\mu$ where $\mu$ is the scale of soft
interaction. Therefore, $\as\Lb Q_s\Rb \,\,\ll\,\,1$ and this fact
allows us to treat this system on solid theoretical basis. On the
other hand, even though the strong coupling $\alpha_S$ becomes small
due to the high density of partons, saturation effects, the fields
interact strongly because of the classical coherence. This leads to a
new regime of QCD with non-linear features which cannot be investigated
in a more traditional perturbative approach. In the framework of the CGC approach the secondary hadrons are
originated from the decay of gluon mini jets with the transverse
momentum approximately equal to the saturation scale $Q_s(x)$~\cite{LRPP}. The first stage of
this process is rather under theoretical control and determines the
main characteristics of the hadron production, especially as far as
energy, rapidity and transverse momentum dependence are concerned. The
jet decay and hadronization unfortunately, could be treated mostly phenomenologically.

The CGC~\cite{LRPP} predicted $7$ TeV data in $pp$ collisions
\cite{cms-rapidity} including (i) multiplicity distribution, (ii) inclusive
charged-hadron transverse-momentum distribution, (iii) the position of
peak in differential yield 4) average transverse momentum of the
produced hadron as a function of energy and hadron multiplicity. The
same model also describes $ep$, $eA$ and $AA$ (at RHIC) data in an
unified fashion supporting the universality of the saturation
physics. It has been shown that the observed ridge phenomenon in $pp$
collisions at the LHC can be also explained by the CGC~\cite{rid}.
There exists some ideas how to simulate the CGC state in $AA$
collisions due to higher density but evidence for the formation of the CGC state (gluon
saturation) in proton-proton interaction will be a triumph of the
high-density QCD and the CGC.

The physics of $AA$ collisions is more complicated compared to $pp(A)$
and $ep(A)$ collisions. The ALICE collaboration has recently released
new data for the multiplicity in central $Pb Pb$ collisions 
at $\sqrtsnn=2.76$ TeV~\cite{alice}. There are some
surprises in the ALICE data: (i) the power-law behavior on energy in
$AA$ is so different from $pp$ collisions which is not very easy to
accommodate within the CGC approach, (ii) the models that describe DIS for
proton, DIS for nucleus, the LHC data for proton and RHIC data
apparently failed to describe the ALICE data with the same accuracy.
It appears apparently to be difficult to describe at the same time,
HERA and RHIC and the new ALICE data for the multiplicity
\cite{talkme}. First notice that, the ALICE $0-5\%$ centrality bin  at $\sqrtsnn=2.76$ TeV
corresponds to $N_{part}=381$ while our approach based on the Glauber
model gives $N_{part}=374$~\cite{LR2}. Therefore, our actual
prediction~\cite{LR2} for the same centrality bin will be higher than what the
ALICE collaboration quoted in their paper~\cite{alice}.  Assuming that
the ALICE data is correct, saturation models gave correct predictions
for multiplicity in $AA$ collisions at the LHC within about less than
$20\%$ error. Indeed, this is not horribly bad given the simplicity of
the approach.  However, this will give rise to several open questions:
(i) what is the role of final-state effects? (ii) how the mini-jet mas
changes with energy/rapidity in a very dense medium?, (iii) what is the
effects of fluctuations and pre-hadronization?. One should also have
in mind that the $k_T$ factorization for $AA$ collisions has not yet
been proven and gluon production in $AA$ collisions is still an open
problem in the CGC. To conclude, the first LHC data for $AA$
collisions has already created much excitement in the heavy ion
community and it opened a fresh and hot debate on how the saturation
physics changes from $ep$, $pp$ and $eA$ to $AA$
collisions.

\tit{Systematic study of inclusive hadron production spectra in collider experiments}

\auth{Andrey Rostovtsev\footnote{In collaboration with A. Bylinkin.} (ITEP, Moscow)}

There exists a large body of experimental data on hadron production
in high energy proton-(anti)proton, photon-proton, photon-photon and heavy ion
collisions.
In the present report the experimentally measured inclusive spectra of
long-lived charged particles produced at central rapidities in the colliding particles
center of mass system are considered.
The analysed published data have been taken with a minimum
bias trigger conditions and at center of mass
energy $(\sqrt{s})$ ranging from 23 to 2360 GeV.

    The charged particle spectra as function  of  transverse
momentum are traditionally approximated using the Tsallis-type (power law) function.
    However, a closer look at the fits to the available data discloses
systematic defects in this approximation.
 It is found, the parameterization
\begin{equation}
\label{eq:exppl}
\frac{d\sigma}{p_T d p_T} = A_e\exp {(-E_T^{kin}/T_e)} +
\frac{A}{(1+\frac{p_T^2}{T^{2}\cdot n})^n}
\end{equation}
is in much better agreement with the data then the Tsallis approximation.
The variables in the equation above are self-explanatory.
    The most surprising feature of the new parameterization~(\ref{eq:exppl})
is a strong correlation between
the parameters $T_e$ and $T$.
Though the physical origin of the observed correlation is not quite
clear, it provides an
additional constraint for the parameterization~(\ref{eq:exppl}) and
therefore reduces a number of free parameters.
Interestingly, a similar combination of the Boltzmann-like and power-law terms is observed in the photon
energy spectra from the sun flares.

The relative contributions of the terms in~(\ref{eq:exppl})
are characterized by a ratio $R$ of the exponential to power law
terms integrated over $p_T^{2}$. 
Interestingly, for $pp$ and $p\overline{p}$ data this ratio $R$ is almost independent of the
collision energy and equals to about $4$, while for $Au Au$ it reaches minimum values 
(about $2$) at medium centralities of heavy ion collisions.
In addition, in the high energy DIS, photoproduction and
$\gamma\gamma$ collisions the power law term of the new proposed
parameterization~(\ref{eq:exppl}) dominates the produced particle
spectra. Thus, only the inclusive spectra of
 charged particles produced in pure baryonic collisions require a
substantial contribution
of the Boltzmann-like exponential term.

Finally, a map of the parameters $T$ and $n$ for proton-(anti)proton, heavy ion,
$\gamma$-proton and $\gamma \gamma$ collision at different energies is drawn.
There are two clearly distinct trends seen on the map. The $pp$ and $p\overline{p}$ 
collision data show an increase of the $T-$parameter and decrease of the
$n$ parameter with collision energy $\sqrt{s}$ increasing.
The second trend, where the values of both parameters the $T$ and $n$
increase, is defined mainly by the RHIC $Au Au$
collision data at $\sqrtsnn=200~GeV$ per nucleon. In this case a
simultaneous increase of
the $T$ and $n$ values corresponds to an increase of the
centrality of heavy ion collisions.
 Surprisingly, the  both trends cross each other at medium
centralities corresponding to the minimum bias $Au Au$ collisions and
$p\overline{p}$ interactions with energy of $\sqrtsnn$=200~GeV.
Naively one could expect the single $p\overline{p}$ interaction has more
similarity to the very peripheral single nucleon-nucleon interactions.
Contrary to that, DIS, $\gamma{p}$ and $\gamma\gamma$ interactions
belong to the second trend and are
located on the
parameter map nearby very peripheral heavy ion interactions at
about the same collision energy per nucleon.
A more extended version of this report can be found in~\cite{BR}.

\tit{Hyperon transverse momentum distributions in $pp$ and $p\bar{p}$ collisions}

\auth{Olga Piskounova (Lebedev Inst., Moscow)}

The analysis of data on hyperon transverse momentum distributions, $dN/dp_T$, that were 
gathered from various experiments (WA89, ISR, STAR, UA1 and CDF) reveals an important 
difference in the dynamics of multiparticle production in proton-proton vs. antiproton-proton 
collisions in the region 
0.3 GeV/c $<p_T<$ 3 GeV/c. Hyperons 
produced with proton beams display a sharp exponential slope at low $p_T$, while those 
produced with antiproton beam do not. Since LHC experiments have proton projectiles, 
the spectra of multiparticle production at the LHC~\cite{cms_mult1} should be ``softer'' in comparison to 
predictions, because the MC predictions were based on Tevatron (antiproton) data. 

\begin{figure}[htbp]
\centering
\epsfig{figure=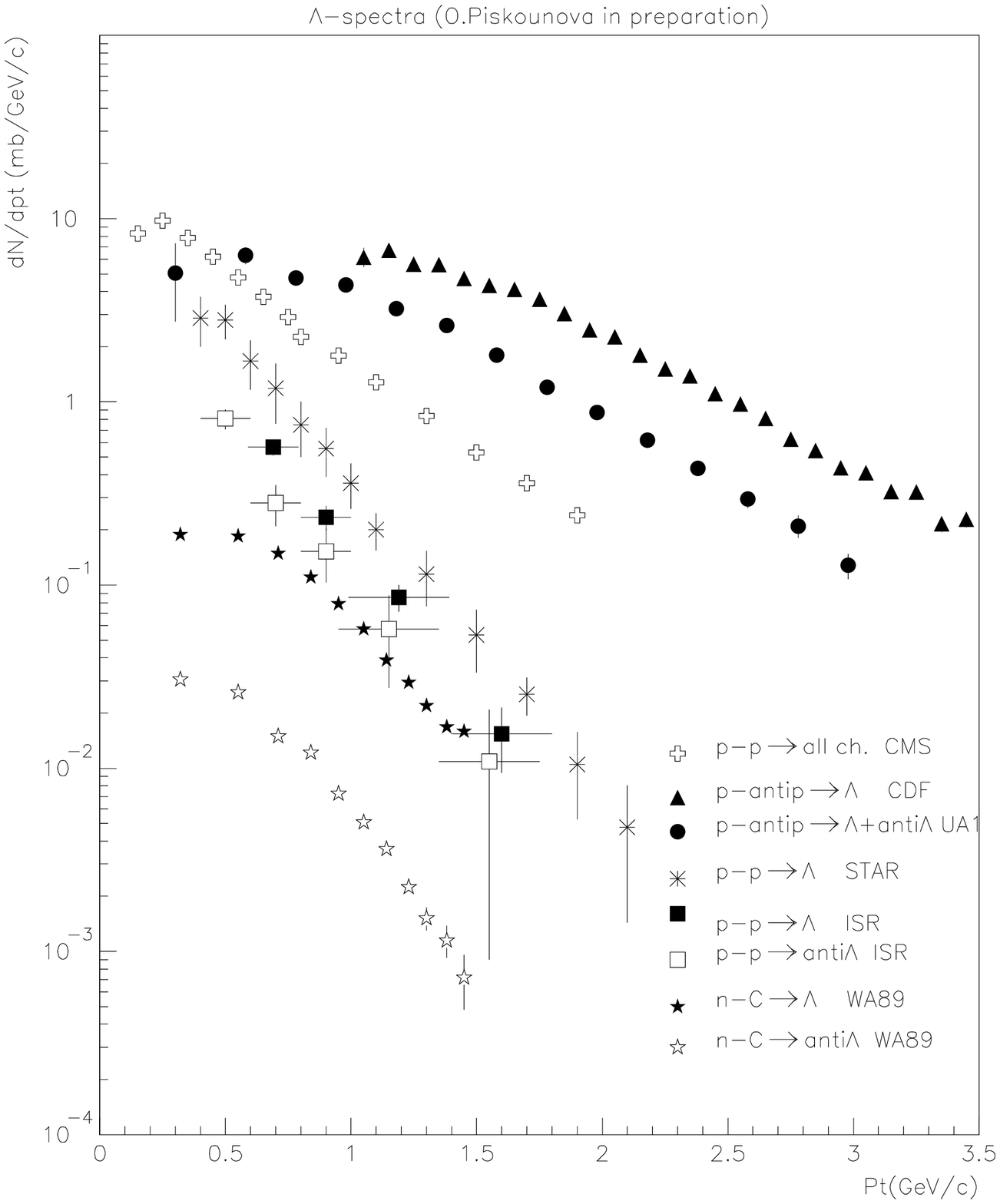,height=7.5cm,width=7cm}
\caption{Strange baryon transverse momentum distributions from different experiments.}
 \label{fig:pisk}
\end{figure}

The available data of many high energy experiments on $p\bar{p}$ collisions~\cite{ua1,cdf} 
as well as on $pp$ collisions of lower energies~\cite{star,isr} and neutron-carbon 
reaction~\cite{wa89} are considered in this article in order to understand the influence 
of quark composition of beam particle on the shape of transverse momentum spectra of 
$\Lambda^0$ hyperon production as at high energy collider experiments as in low energy 
fixed-target experiment (Fig. \ref{fig:pisk}).
The difference in $p_T$ spectra of $\Lambda^0$'s produced in high energy $pp$ and $p\bar{p}$ 
collisions can not be explained in the QCD theoretical models, because at the collider energies 
both interactions should give the multiple particle production due to Pomeron (or multi-Pomeron) 
exchange. 
The total cross section and the spectra in $pp$  and $p\bar{p}$ 
collisions are to be similar because of the Pomeron exchange between two interacting hadrons
that should not be sensitive to the quark contents of colliding beams at high energies. 
 
Unfortunately, this difference was not studied enough at ISR, where both projectiles were
available. The important fact is that the latest collider experiments 
were carried out with antiproton beams. It was mistake to suggest that $pp$ and $p\bar{p}$ at
high energy are giving the similar transverse momentum distributions. The spectra of hyperons 
that are produced with proton beam have a sharp exponential slope at low $p_T$, while the 
spectra with antiproton beam have not. The highest energy experiments (UA1 and Tevatron) shows 
harder $p_T$ spectra, that is not only the result of growing energy -- it is 
the result of different form of transverse momentum distributions in different reactions. 

From the point of view of the Quark-Gluon String Model ({\sc qgsm})~\cite{qgsm}, the most important contribution 
to particle production spectra in antiproton-proton reactions is due to antidiquark-diquark 
string fragmentation.
Baryon hadroproduction spectra are sensitive to quark-diquark structure of interacting hadrons 
as well as to the energy splitting between these components.
Asymmetric reactions may provide us with a new ``stereoscopic'' view on the hadroproduction mechanism.
Measurements of $p_T$ spectra in antiproton-proton interactions at a variety of energies can 
thus constrain the contribution from the fragmentation of antidiquark-diquark string. This 
study may have impact not only on the interpretation of LHC results, but also on cosmic ray 
physics and astrophysics, where matter-antimatter asymmetry is being studied.

\tit{\dpmjet--III and $p p$ data from the LHC}

\auth{Johannes Ranft (Siegen University) }

Monte Carlo codes based on the two--component Dual Parton Model
(soft hadronic chains and hard hadronic collisions) are
available since 10--15 years: \phojet\ for hadron-hadron ($h$--$h$) 
and photon-hadron ($\gamma$--$h$) collisions~\cite{phojet-a} and \dpmjet-III based on
\phojet\ for hadron-nucleus ($h$--$A$) and nucleus-nucleus ($A$--$A$) collisions~\cite{dpmjet2}.

At the LHC particle production in $p p$ collisions was measured
by the three Collaborations CMS, ALICE and ATLAS. Here we
compare with measured pseudorapidity distributions at 900, 2360 and 7000
GeV c.m. energy, with $p_T$ distributions of charged
hadrons at the same energies and with $\bar p$ to $p$ ratios.
The problem to be solved at the beginning was, that \dpmjet-III
and all other event generators did predict pseudorapidity
distributions rising slower with energy than the LHC data.
To solve this problem we had to redetermine the parameters of
\dpmjet-III in such a way, that agreement with the data is
achieved. We can present at the moment only one preliminary solution.
In this solution we introduce an energy dependence in two of the
\dpmjet-III parameters. We call this a preliminary solution,
since we think that also the new parameters should not depend on
the energy. We are confident, that we will find soon such a
solution, but at the moment all solutions with energy
independent parameters have still problems. 

We were able to present at the meeting charged pseudorapidity
distributions which agree perfectly with the CMS data.
We found also central antiproton to proton ratios in agreement
with the ALICE data. The $p_T$ distributions of \dpmjet-III agree
with the CMS distributions. But one weakness of the $p_T$
comparisons is, that at present the average $p_T$ values at
energies lower than the LHC energies are slightly higher than
the data, this is also one problem which has to be solved.
Finally, we find perfect agreement of charged hadron
multiplicity distributions comparing \dpmjet-III with the
ALICE data.

\newpage

\section{Cosmic-rays at Ultra-High Energies: Experiments}

\tit{Results from the Pierre Auger Observatory}

\auth{Lorenzo Cazon\footnote{On behalf of the PAO Collaboration. Support 
FCT-Portugal (CERN/FP/109286/2009) and ECT* are acknowledged.} (LIP, Lisbon)}

The Pierre Auger Observatory is the largest cosmic ray (CR) observatory on Earth, covering 
3000 km$^2$ of the high plateau in the Argentinian region of Pampa Amarilla~\cite{Auger}. 
It detects CR air showers in two complementary ways: an array of water-\v{C}erenkov tanks samples 
the secondary particles at ground and fluorescence telescopes observe the longitudinal development 
of the electromagnetic cascade~\cite{SDFDHY}. 

An energy spectrum has been recently published~\cite{spectrum} covering the energy range from $10^{18}$~eV 
to above $10^{20}$~eV. The dominant systematic uncertainty stems from the overall energy scale, 
and is estimated to be 22\%. The position of the ankle at $\log_{10}(E_{ankle}/eV) = 18.61 \pm  0.01$ 
and a flux suppression above $\log_{10}(E_{1/2}/eV) = 19.61 \pm  0.03$ have been determined. 
The suppression is similar to what is expected from the GZK effect for protons or nuclei as 
heavy as iron, but could also be related to a change of the injection spectrum at the sources.

\begin{figure}[htbp]
\centering
\includegraphics[width=13.5cm]{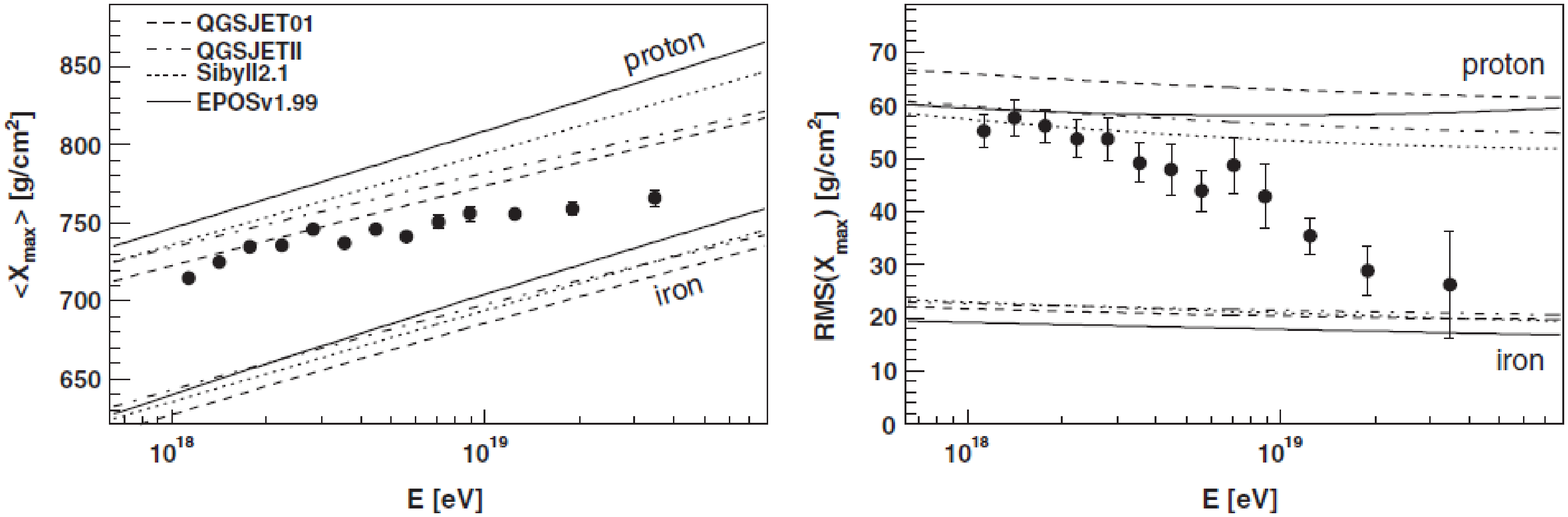}
 \vskip -0.2cm
\caption{$\mean{X_{max}}$ and RMS($X_{max}$) measured by Auger compared to different hadronic interaction models.
\label{figXmax}}
\end{figure}

Analyses of both the mean values and the fluctuations of the  shower maximum $\mean{X_{max}}$ 
(plotted in Fig.~\ref{figXmax}, in comparison with the prediction of different hadronic interaction models) 
reveal a change in the energy dependence of the composition around the ankle and a gradual increase 
of the average mass of cosmic rays with energy, provided that there are not significant changes in the 
properties of the hadronic interactions at ultrahigh energies.  More details can be found in~\cite{Xmax}.

A detailed comparison of the muonic and electromagnetic content of the air showers far from the core shows 
an excess in the number of muons of $\sim50\%$ compared to protons simulated with \qgsjet-II ($\sim 30\%$ 
excess compared to iron)~\cite{muons}. This analysis also suggests a larger energy scale (27\%) with respect 
to the fluorescence detector, being still compatible within the systematic uncertainties.

Showers initiated by photons and neutrinos have distinct signatures compared to showers initiated by protons 
or other nuclei, being possible to discriminate them~\cite{neutrinos,photons} and set bounds on their fluxes.
The neutrino and photon bounds already exclude `top-down' models for the production of ultra-high energy 
cosmic rays, favoring acceleration in astrophysical scenarios. 

In 2007, it was shown that the arrival directions of  CRs with energy in excess of 55 EeV correlated with 
the positions of nearby AGN~\cite{science}. As further data has been added, the degree of correlation has 
decreased from $69^{+11}_{-13}$\% to $38^{+7}_{-6}$\%, to be compared with the 21\% expected if the flux were 
isotropic~\cite{UpdateAGN}. The region of the sky with the largest observed excess with respect to isotropy 
corresponds to Cen A, which is the closest AGN.  Nevertheless, at present there are multiple astrophysical 
models of anisotropy which are fully consistent with the observed distribution of arrival directions.
While a correlation of arrival directions with  nearby matter on small angular scales is plausible for protons
above 55 EeV, it is puzzling if the CRs are heavy nuclei as suggested by the $X_{max}$ measurements, since 
they are expected to undergo large deflections due to the galactic and extragalactic magnetic fields. 
Definitive conclusions must await additional data.


\tit{HiRes results: The final word (almost)}

\auth{Douglas R. Bergman\footnote{On behalf of the HiRes collaboration.} (Univ. of Utah)}

\newcommand{\xmax}{\ensuremath{X_{\rm max}}}

The High Resolution Fly's Eye (HiRes) experiment was a ultra-high
energy cosmic ray detector operating in Utah from 1997-2006.  It used
the fluorescence technique to detect cosmic ray air showers in stereo
using two detector sites.

HiRes was the first experiment to observe the GZK cutoff.  We observed
the break with a significance of greater than 5$\sigma$ in monocular
mode~\cite{Abbasi-2007-PRL-100-101101} and confirmed that discovery
in stereo mode~\cite{Abbasi-2009-APP-32-53}.  The position of the
break is consistent with what is expected for extragalactic protons as
measured with Beresinki's $E_{1/2}$ test
\cite{Berezinsky-Grigoreva-1988-AA-199-1}. 

HiRes has made a direct measurement of cosmic ray composition at
energies above $10^{18.2}$ eV using both the average \xmax\ technique
and by measuring the width of the \xmax\ distribution
\cite{Abbasi-2010-PRL-104-161101}.  Both show a composition
consistent with pure protons.  We note that great care must be taken
to account for both trigger and reconstruction biases for the mean
\xmax\ measurement. We also note that the RMS is a biased estimator of
the width of a non-Gaussian distribution; we use a Gaussian fit to the
central part of the distribution (within twice the RMS of the mean) as
a better estimator of the width.  HiRes also measured the average
width of individual showers.  This also agrees with a protonic
composition.

HiRes has made a number of searches for anisotropy in the arrival
directions of cosmic rays.  We observed none: no correlation with
AGN's~\cite{Abbasi-2008-APP-30-175}, no correlation with the local
large scale structure as determined by galaxy surveys
\cite{Abbasi-2010-arXiv:1002.1444}.

\tit{Telescope-Array results}

\auth{Nobuyuki Sakurai\footnote{On behalf of the Telescope-Array collaboration.} (Osaka City Univ.)}

Telescope Array (TA) is a largest ultra high energy cosmic ray detector in northern hemisphere.
Operation was started about 3 years ago, and the gathered data have delivered the interesting 
results on the extremely high energy cosmic rays.
The detector consists of 507 plastic scintillation detectors (SD) which cover the ground area of 
680 km${}^{2}$ in 1.2 km mesh and 3 fluorescence telescope stations (FD) which surround the 
scintillator detector array and look inward.
Each detector of SD has 2 layers of plastic scintillator plate of 3~m${}^{2}$ area and 1.2 cm thickness.
For FD, Telescope Array adopts two different types of telescopes.
Two FD stations, which are called as BRM-FD and LR-FD, are newly developed for Telescope Array.
One FD station which is called as MD-FD consists of the telescopes which had been operated as Hires-I detector.\\

We analyze energy spectra of ultra high energy cosmic ray using 3 different data sets: FD-mono, Hybrid and SD.
In the FD-mono analysis, MD-FD data is analyzed by the same program as HiRes-I.
And we obtained energy spectrum which is consistent with HiRes result. 
This means that HiRes-I detectors which was moved to Telescope Array site have reproduced HiRes result.
The Hybrid energy spectrum is obtained from the analysis which use both of FD data and SD data in order 
to improve the geometrical reconstruction, although its energy is reconstructed as same as FD analysis.
The energy spectrum obtained by the hybrid analysis is consistent with FD-mono result.
TA SD energy spectrum is also consistent with spectra from both of FD-mono analysis and the hybrid analysis.
In analysis of SD data, SD energy scale is scaled so as to agree with FD energy scale using the result of hybrid analysis. 
Primary composition is studied using the shower maximum (X${}_{max}$) observed by FD stereo data.
In the energy region of $10^{18.6} \sim$ $10^{19.3}$~eV, the averaged X${}_{max}$ is consistent 
with the proton primary hypothesis.
Arrival direction study shows no correlation with Active Galactic Nuclei so far.
The existence of ultra high energy photon is studied using the shower front curvature observed by SD, 
but no candidate is found in data.
Electron LINAC (ELS) is installed in front of BRM-FD to calibrate using the electron beam in this year.
The ELS calibration is expected to improve the systematic error of FD drastically.

\tit{ARGO-YBJ results}

\auth{Ivan De Mitri\footnote{On behalf of the ARGO-YBJ collaboration.} (INFN, Lecce)}

Cosmic ray physics in the $10^{12}-10^{15}\,$eV primary energy range is among the main scientific goals of
the ARGO-YBJ experiment~\cite{bacci2002,demitri2007cris}. The detector, located at the Cosmic Ray Observatory of
Yangbajing (Tibet, P.R. of China) at 4300~m a.s.l., is a full coverage Extensive Air Shower array consisting
of a carpet of Resistive Plate Chambers of about 6000$\,$m$^2$. The apparatus layout,
performance and location offer a unique possibility to make a deep study of several characteristics
of the hadronic component of the cosmic ray flux in an energy window marked by the transition
from direct to indirect measurements.
In this short summary we will focus on hadronic interaction studies that are being performed within the
experiment in the primary energy range going from 1 TeV to 1 PeV. 

The proton-air cross section has been measured.
The total proton-proton cross section has then been estimated at center of mass energies
between 70 and 500~GeV, where no accelerator data are currently available. 
Other hadronic interaction studies can be performed by exploiting the detector capability to
have very detailed information on the shower front space-time structure and the lateral 
distribution function by also using the analog readout of the RPC's.
Because of lack of space, here we will report on the $p$-air and $p p$ cross section measurement only.

The measurement is based on the shower flux attenuation for different 
zenith angles, i.e. atmospheric depths~\cite{xsecpaper}. 
The detector location (i.e. small atmospheric depth) and features (full coverage, angular resolution, 
fine granularity, etc.) ensure the capability of reconstructing showers in a 
very detailed way. These features have been used to fix the energy ranges and to constrain the shower ages.
In particular, different hit (i.e. strip) multiplicity intervals have been used to select showers 
corresponding to different primary energies.
At the same time the information on particle density, lateral profile and shower front extension have 
been used to select showers having their maximum development within a 
given distance/grammage $X_{dm}$ from the detection level.
This made possible the unbiased observation of the expected exponential falling of shower intensities as
a function of the atmospheric depth through the $\sec\theta$ distribution.
After the event selection, the fit to this distribution with an exponential law gives the slope 
value $\alpha$, connected to the characteristic length $\Lambda$ through the 
relation $\alpha = h_0 /  \Lambda$. That is:
\begin{equation}
  I(\theta) = A(\theta) \, I(\footnotesize{\theta=0}) \,e^{-\alpha \, (\sec\theta - 1)} 
  \label{eq:sectheta}
\end{equation}
where $A(\theta)$ accounts for the geometrical acceptance of each angular bin. 
The parameter $\Lambda$ is connected to the proton interaction length by the relation $\Lambda = k \lambda_{int}$, 
where $k$ depends on hadronic interactions and on the shower development in the atmosphere and its 
fluctuations~\cite{ulrich2007}.
The actual value of $k$ must be evaluated by a full MC simulation and it depends also on
the experimental approach, the primary energy range and on the detector response. 
The $p$-air {\it production}~\cite{production} cross section is then obtained from the relation:
$\sigma_{p-air}\,$(mb) $= 2.41 \times 10^{4} / \lambda_{int}\,$(g/cm$^2$), while several 
theoretical approaches can be used to get the corresponding $p p$ total cross section $\sigma_{pp}$ 
\cite{pairtopp}.
 
\begin{figure}[htbp]
\centering
\includegraphics [width=0.65\textwidth,height=0.35\textwidth]{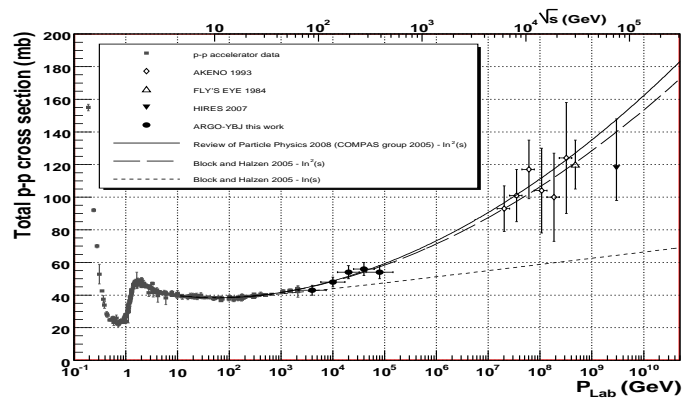}
 \vskip -0.2cm
  \caption{The $p p$ total cross section obtained by ARGO-YBJ, 
           together with results published by other cosmic ray and accelerator experiments~\cite{xsecpaper}.}
  \label{fig:sigma_pp}
\end{figure}

Results are shown in Fig.~\ref{fig:sigma_pp}.
As can be seen, ARGO-YBJ data lie in an energy region not yet reached by $p p$ colliders (and 
still unexplored by $p \bar{\rm p}$ experiments), favouring 
the asymptotic $ln^2(s)$ increase of total hadronic
cross sections as obtained in~\cite{blockhalzen2005} from a global analysis of accelerator data.

\tit{KASCADE-Grande results}

\auth{Paul Doll\footnote{On behalf of the KASCADE-Grande collaboration.} (KIT, Karlsruhe)}

Testing of hadronic interaction models \qgsjet-II-2 and \epos\ 1.99 implemented in the \corsika\
program have been performed with KASCADE-Grande air shower data in the energy range of 10$^{16}$ 
to 10$^{18}$ eV~\cite{PD1,PD2}. From the muon density investigations, the \epos\ 1.99 model indicates that light 
abundances of primary cosmic ray particles would be needed to fit the data. On the other hand, 
the \qgsjet-II-2 model describes the data with an intermediate primary abundance between proton 
and iron nuclei. The reconstructed all-particle energy spectra are presented by using the 
hadronic interaction models \qgsjet-II-2 and \epos\ 1.99. The resulting spectra show that the 
interpretation of the KASCADE-Grande data with \epos\ 1.99 leads to significantly higher flux as 
compared to the \qgsjet-II-2 result. More detailed investigations of \epos\ 1.99 are still in work.\\

\clearpage

\tit{IceCube results}

\auth{Lisa Gerhardt\footnote{On behalf of the IceCube collaboration.} (LBL, Berkeley)}

High energy neutrinos offer a unique view of distant, energetic astrophysical objects, 
as they are neither bent by ambient magnetic fields nor absorbed by the interstellar
 medium. Possible sources of neutrinos include active galactic nuclei, gamma ray bursts, 
the highest energy cosmic rays, and interactions of exotic objects. The IceCube neutrino 
detector uses the ice at the South Pole as a \v{C}erenkov medium for the detection of 
high energy neutrinos. It is composed of an in-ice, three-dimensional array of photomultiplier 
tubes~\cite{pmt} and a surface air shower array. 
Construction of the IceCube detector began in 2005 and was finished in 2010, bringing the detector 
to its full cubic kilometer size. Using data from the partially constructed detector, the 
IceCube Collaboration has searched for point sources of neutrinos~\cite{ps1,ps2} and found
results consistent with the expectation from the background of atmospheric neutrinos. 
It has set stringent upper limits on the diffuse fluxes of extremely high energy 
neutrinos~\cite{ehe} and on the flux of neutrinos in coincidence with Gamma-Ray Bursts~\cite{grb} 
and has set limits on the accumulation of dark matter in the Sun~\cite{wimp1,wimp2}. 
It has measured the flux of atmospheric neutrinos up to 400 TeV~\cite{atm} and the anisotropy 
of the arrival directions of cosmic rays with a median energy of 20~TeV~\cite{aniso}. 
IceCube data collection continues.

\newpage

\section{Cosmic-rays at Ultra-High Energies: Theory}

\tit{Open problems in cosmic ray physics and the importance of understanding hadronic interactions}

\auth{Ralph Engel (KIT, Karlsruhe)}

\begin{figure}[htb!]
\centering
\includegraphics[width=0.60\textwidth]{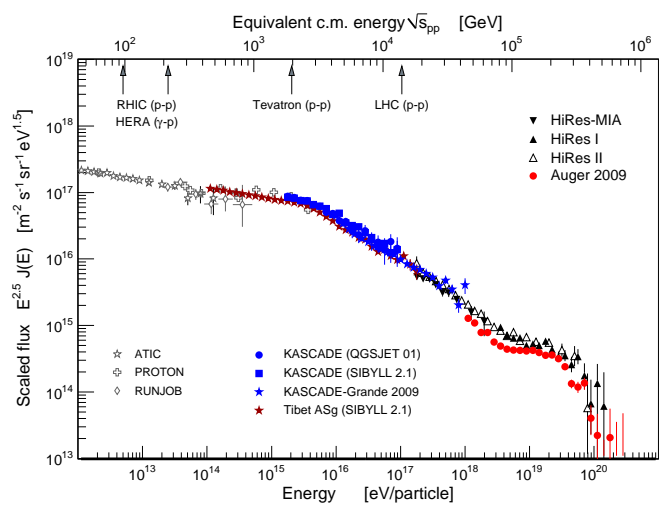}
 \vskip -0.5cm
\caption{\label{fig:CR-flux-all} The all-particle flux of cosmic rays. For references to the data see~\cite{Bluemer:2009zf}.}
\end{figure}

In Fig.~\ref{fig:CR-flux-all}, a compilation of measurements of the
all-particle spectrum of cosmic rays is shown (from
\cite{Bluemer:2009zf}, updated). The most striking features are the
knee at about $3\times 10^{15}$ eV, the ankle between
$10^{18} - 10^{19}$eV, and the suppression of the flux at the
very highest energies. Understanding the origin of these
characteristic breaks in the power-law of the flux is key to
identifying the galactic and extragalactic sources of cosmic rays and
the corresponding particle acceleration and propagation mechanisms.

For example, if the knee in the flux stems from features of diffusive
shock acceleration processes it is expected that the fluxes of
individual elements will exhibit knee-like features shifted in energy
according to the magnetic rigidity $\sim E/Z$. Similarly, particle
physics scenarios lead to the prediction of a scaling with mass number
$\sim E/A$. Many other models have been developed for the knee, for
example~\cite{Stanev:1993tx,Erlykin:2010yy} -- for a review, see
\cite{Hoerandel:2002yg}. Similarly, there are several
interpretations of the ankle. It seems natural to assign this feature
to the transition between galactic and extragalactic cosmic ray
sources~\cite{Hillas:2005cs}. Alternatively, in the dip model the
ankle is the imprint of $e^+ e^-$ pair production in extragalactic
propagation~\cite{Berezinsky:2002nc}, requiring a proton-dominated
composition.

All these model scenarios differ in the predicted composition, making
the accurate measurement of the mass composition and its evolution
with energy a prerequisite to making progress in the field. Currently
the largest uncertainty in the composition interpretation of air
shower data is related to the description of hadronic interactions
that has to be done with phenomenological models~\cite{Knapp:2002vs}.
Measurement of multiparticle production at HERA, RHIC and LHC and
using the data for refining existing interaction models will allow us to make
significant progress in reducing this uncertainty.\\

\clearpage

\tit{UHECRs and hadronic interactions}

\auth{Paolo Lipari (Roma)}

The interpretation of the data on the Ultra High Energy Cosmic Rays
requires an understanding of the development of hadronic showers
and therefore a sufficiently accurate description of the properties
of hadronic interactions. 
Since the energy spectrum of CR extends up to 
$E \sim 10^{20}$~eV, this requires an extrapolation to 
the c.m. energy of $\sqrt{s} \sim 430$~TeV. 

The observations of fluorescence light, pioneered by the 
Fly's Eye collaboration, allows one to estimate 
the energy of a CR particle in a quasi--calorimetric way,
with only little model dependence. The study of the shape 
of the longitudinal shower development allows one in principle to determine
the mass number $A$ of the primary particle, but is also strongly
dependent from the hadronic interaction properties.
The position of the maximum of the 
shower longitudinal development $X_{\rm max}$ is a good indicator 
of $A$. 
The average $X_{\rm max}$  for a primary particle
of energy $E$ and mass $A$, in reasonably good approximation, is given by:
\begin{equation}
\langle X_{\rm max}^{(A)} (E) \rangle \simeq
\left \langle X_{\rm max}^{(p)} \left ( \frac{E}{A} \right )
\right \rangle 
 \simeq X_0 + D_p \; \log \left ( \frac{E}{A} \right ) 
= 
X_0 + D_p \;\log E - D_p \log A 
\label{eq:xmax-dev}
\end{equation}
(the quantity $D_p$ is known as the ``elongation rate'').
This equation allows one to estimate the average mass (or 
$\langle \log A \rangle$) and the 
evolution with energy of the composition of UHECR, that is $d\langle \log A \rangle/dE$,
if one has a good theoretical control of the model dependent
quantities $X_0$ and $D_p$. 
It is therefore necessary to estimate
the systematic uncertainty in the 
calculation of these quantities.
It is also interesting to
discuss what properties of the hadronic
interactions determine $X_0$ and $D_p$.
The main contributions come from the 
hadron interactions lengths 
(including the interaction lengths of mesons) and the inclusive 
energy spectrum of secondary particle in the projectile
fragmentation region.
The  results  that can be obtained at LHC  with a precision measurement
of the $pp$ cross section and  of particle  production properties
will be important in constraining  the models.  

It is interesting to note that it is in principle possible to determine
the composition of CR without any use of shower development, and therefore
use the  CR measurements to obtain information about the  properties
of hadronic  interactions.
One possibility is the observation of the imprints of energy losses
on the observed energy spectrum
(since the kinematical thresholds for the processes of $e^+e^-$ pairs
and pion production are $A$ dependent), and the 
estimate of the deviation due to astronomical magnetic fields
(that depends on $Z$). This could in principle allow one to
determine properties of hadronic interactions from CR observations.
The observations are at this point inconclusive.
The energy scale of the HiRes detector is consistent with the 
hypothesis of attributing the ``ankle'' spectral feature, and the high 
energy suppression with energy loss imprints on a smooth 
spectrum strongly dominated by protons, but this interpretation
is not consistent with the energy scale of the AUGER experiment.
The AUGER  collaboration has  observed a correlation
between the direction of the highest energy 
particles  with potential  extragalactic  sources
but the interpretation of the results remains ambiguous.

The study of the width of the distribution 
in the measurement of $X_{\rm max}$
also allows one to estimate the mass composition of CR, since 
the development of large $A$ primary particle has smaller fluctuations.
The results of the AUGER experiment 
suggest that the highest energy particles are large
$A$ nuclei, however these results are not confirmed by
results of the HiRes and Telescope Array collaborations.

\clearpage

\tit{On the relation between air shower predictions and features of hadronic interactions}

\auth{Ralf Ulrich (KIT, Karlsruhe)}

The nature of cosmic ray particles at the very highest energies is
still not understood. Even with experiments like the Pierre Auger
Observatory~\cite{Abraham:2004dt}, HiRes~\cite{Abbasi:2004nz} and
Telescope Array~\cite{Kawai:2008zza} delivering large quantities of
high quality data, it is not straightforward to interpret these
data. The fundamental problem is that the observations are very
indirect: Only extensive air shower cascades, which are initiated by
the cosmic ray primaries, are observed. For an accurate analysis of
these air shower data a detailed understanding of interactions in the
cascades is required. The particle production characteristics that are
important in the context of air showers are interactions at energies
of up to $\sqrt{s}\sim$~350~TeV and particle production in the
very forward direction ($\eta>6$). Particle production
characteristics in this phase space have a strong impact on the
modelling of the evolution of air shower
cascades~\cite{Ulrich:2010rg}. 

\begin{figure}[htbp!]
  \centering
  \includegraphics[width=.6\textwidth]{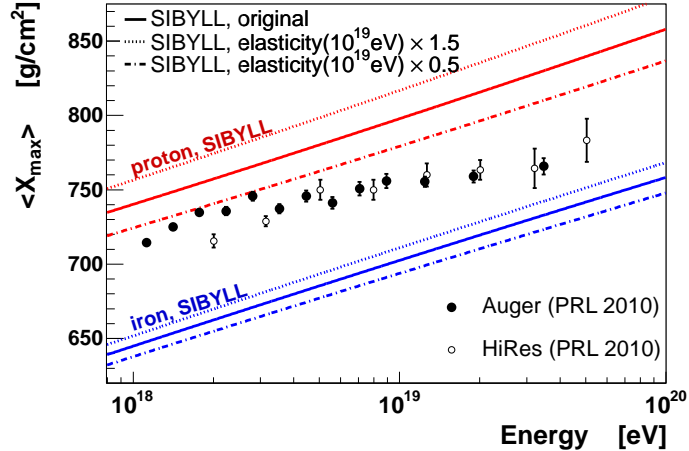}
  \vspace*{-.2cm}
  \caption{\label{fig:xmax}Measured average depth of the shower
    maximum~\cite{Abraham:2010yv,Abbasi:2009nf} of UHE 
    cosmic rays compared to \sibyll\ predictions with a modified
    extrapolation of the elasticity to ultra-high energies.}
\end{figure}

In Figure~\ref{fig:xmax} we demonstrate
this at the example of the elasticity, $k_{\rm ela}=E_{\rm max}/E_{\rm tot}$, 
of interactions. It is shown that the interpretation of
cosmic ray data relies on the detailed understanding of hadronic interaction
physics. The LHC has the potential to significantly reduce the
uncertainties because of two reasons: Firstly, it operates at
energies that are already very relevant in terms of cosmic ray
observations and, secondly, it has significant capabilities to study
the forward phase space of multiparticle production that are relevant for the
air shower modelling.

\tit{The event generator \sibyll~2.1}

\auth{Ralph Engel\footnote{In collaboration with Eun-Joo Ahn, Thomas K.~Gaisser, 
Paolo Lipari, and Todor Stanev.} (KIT, Karlsruhe)}

\sibyll\ is an event generator optimized for simulating high energy
interactions needed for the description of extensive air showers and
for the calculation of inclusive muon and neutrino fluxes. In
comparison to other event generators, a rather basic and
straightforward model for the
description of multiparticle production is implemented. The initial
version of the model~\cite{Fletcher:1994bd} was
upgraded in 2000 to include post-HERA
parton density parametrizations and to introduce multiple soft
interactions and a better treatment of
diffraction dissociation~\cite{Ahn:2009wx}.

Proton/pion/kaon-proton collisions are simulated in terms of
multiple partonic soft and hard interactions, where each such partonic
interaction leads to two QCD color strings, which subsequently
fragment into hadrons. Diffraction dissociation is implemented as
two-channel model of excited states for projectile and target
particles, similar to the Good-Walker model~\cite{Good:1960ba} of diffraction
dissociation. The minijet cross section is calculated within the
QCD-improved parton model using the parton density parametrization of
Gl\"uck, Reya and Vogt~\cite{Gluck:1994uf}. Saturation is accounted for by introducing an
energy-dependent transverse momentum cutoff for distinguishing soft
and hard interactions~\cite{Gribov:1984tu}. 

\begin{figure}[htb!]
\centerline{
 \includegraphics[width=0.3688\textwidth]{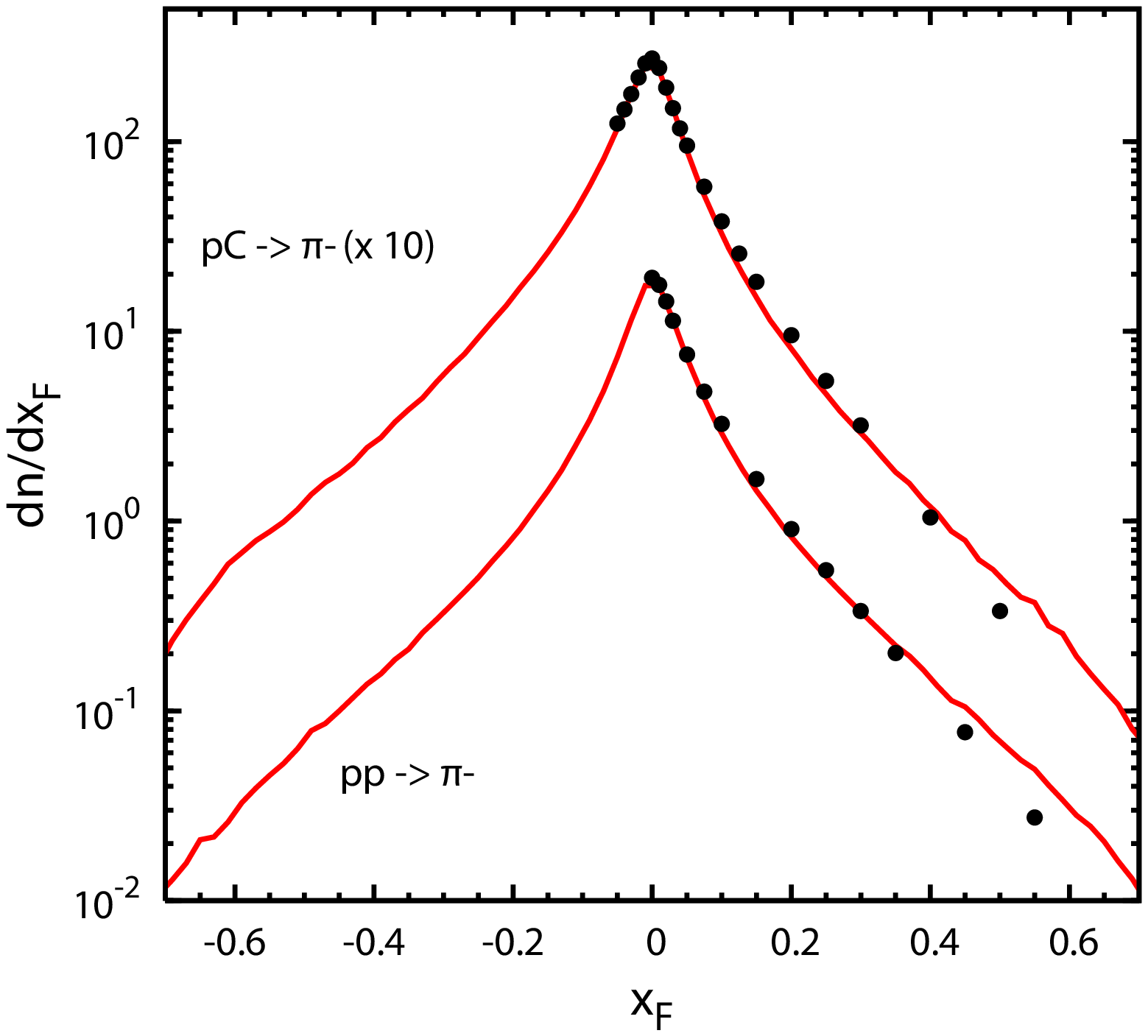}
\hspace*{2cm}
 \includegraphics[width=0.4\textwidth]{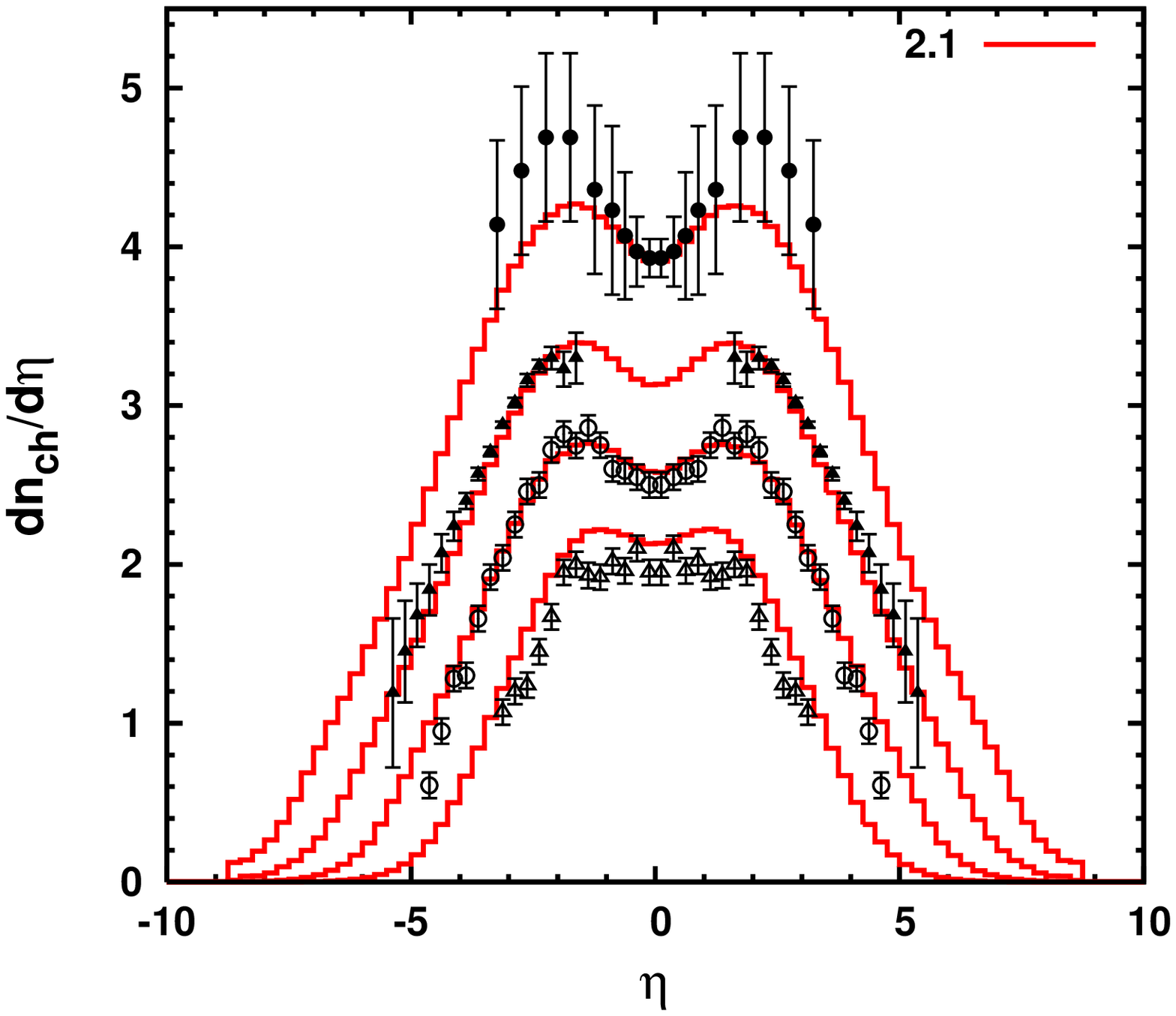}
}
 \vskip -0.2cm
\caption{
\label{fig:SIBYLL}
Comparison of \sibyll\ predictions with fixed target data from NA49~\cite{Alt:2005zq} at $E_{\rm lab} =
158$\,GeV (left) and different data sets from collider 
experiments (right) $\sqrt{s} = 54 - 1800$\,GeV~\cite{Ahn:2009wx}.
} 
\end{figure}

The scattering of hadrons off nuclei is modeled using the Glauber
approximation. Realistic shell model distributions for the nucleon
densities in the different nuclei are implemented~\cite{Glauber:1970jm}. The
semi-superposition model is used for the description of the interaction of
nuclei as beam particles with air~\cite{Engel:1992vf}.
The MC results are compared to fixed-target and collider data in Fig.~\ref{fig:SIBYLL}.

\tit{Pomeron calculus: cross sections, diffraction and MC simulation}

\auth{Sergey Ostapchenko (NTNU, Trondheim)}

Monte Carlo (MC) generators of hadronic interactions are standard
tools for data analysis in high energy collider and cosmic ray (CR)
fields. A general framework for developing such generators is provided
by the Reggeon Field Theory ~\cite{gri68}. The key technique
is the one proposed by Abramovsky, Gribov and Kancheli (AGK)~\cite{agk},
which allows one to relate partial cross sections for various configurations
of hadronic final states to certain unitarity cuts of elastic scattering
diagrams. The procedure generally requires the knowledge of the amplitude
for an ``elementary'' elastic rescattering process (``Pomeron exchange'')
which corresponds to an underlying microscopic parton cascade, of
the vertices for Pomeron-hadron coupling, and of the vertices for
Pomeron-Pomeron ($\Pom\Pom$) interactions, if non-linear interaction effects are
to be taken into account. In principle, all the three ingredients
may be specified using different approaches, 
ranging from purely phenomenological
parametrizations to BFKL-based treatment. In practice, one usually
considers ``soft'' and ``semihard'' contributions to the elastic rescattering
process, depending on whether it is dominated by a purely nonperturbative
soft parton cascade or parton evolution extended to moderately
large virtualities $|q^{2}|$. While soft parton evolution is described
by phenomenological parametrizations (e.g., as soft Pomeron emission),
its extension to higher $|q^{2}|$ is typically treated within the
DGLAP formalism.

It is worth stressing that the starting point for a development of
a MC generator is the derivation of the complete set of partial cross
sections for various hadronic final states. The latter are defined
by the number of elementary particle production processes (``cut Pomerons'')
and by a particular arrangement of those cut Pomerons in rapidity-impact
parameter space. Thus, calculations of partial cross sections involve
full resummation of virtual (elastic) rescatterings which are described
as uncut Pomeron exchanges.
Such an analysis becomes especially simple when using eikonal
vertices for Pomeron-hadron coupling and neglecting Pomeron-Pomeron
interactions, which leads to the standard eikonal description of the
elastic scattering amplitude and to simple expressions for partial
cross sections. In particular, one arrives to the Poisson distribution
for a number of elementary production processes for a given impact
parameter.

However, the necessity to describe non-linear corrections to the interaction
dynamics, related to parton shadowing and saturation, forces one to
take Pomeron-Pomeron interactions into account, which significantly
complicates the formalism. Indeed, with the energy increasing, so-called
enhanced ($\Pom\Pom$ interaction) graphs of more and more complicated
topologies start to contribute significantly to the scattering amplitude
and to partial cross sections for particular hadronic final states.
Thus, dealing with enhanced diagrams, all order resummation of the
corresponding contributions is a must, both for elastic scattering
diagrams and for the cut diagrams representing particular inelastic
processes. Secondly, it is far non-trivial to split the complete set
of cut enhanced diagrams into separate classes characterized by positively-defined
contributions which could be interpreted probabilistically and employed
in a MC simulation procedure. While the first problem has been addressed
in~\cite{ost06,ost06a,ost10}, the MC implementation of the
approach has been discussed in~\cite{ost10a}. Let us briefly list
the main results of the analysis of Refs.~\cite{ost06,ost06a,ost10,ost10a}. 
\begin{itemize}
\item Non-linear contributions to the interaction dynamics are not 
small corrections, but 
dominate the high-energy behavior
of hadronic cross sections and particle production~\cite{ost06,ost06a,ost10a}. 
\item Unlike inclusive particle (jet) cross sections, partial cross sections
of hadronic final states which involve high transverse momentum jets
can not be expressed via universal parton distribution functions (PDFs)
measured in DIS experiments; rather they depend on ``reaction-dependent
PDFs'' which involve parton rescattering on both the parent
(e.g., projectile) hadron and the partner (here, target) hadron~\cite{ost06a}.
In other words, collinear QCD factorization is inapplicable to
 exclusive hadronic final states. 
\item An analysis of partial contributions of different classes of enhanced
diagrams has shown that neither a resummation of contributions of
``fan''-like graphs or of a more general class of ``net''-like enhanced
graphs nor of the ones of ``Pomeron loop'' diagrams alone is sufficient
for a correct description of hadronic cross sections in the high-energy
limit~\cite{ost10}. Instead, both classes of diagrams have to be
 taken into account.
\item Calculations of diffractive cross sections require a proper resummation
of absorptive corrections to the contribution of diffractively cut sub-graphs,
which seriously reduce the probability for a rapidity gap survival,
in addition to the usual (eikonal) rapidity gap suppression (RGS) factor 
\cite{ost10}. In particular, restricting oneself with the simplest 
triple-Pomeron contribution to single high mass diffraction cross section,
one arrives to a contradiction with $s$-channel unitarity, even if the
eikonal RGS factor is included.
\item The complete set of unitarity cuts of elastic scattering diagrams
can be re-partitioned into a number of positively-defined contributions
which define partial cross sections for certain ``macro-configurations''
of the interaction ($s$-channel unitarity)~\cite{ost10a}. For each
of those macro-configurations, the pattern of secondary particle production
can be reconstructed in an iterative fashion using a
MC procedure.
\end{itemize}

One has to mention, however, that 
the approach of~\cite{ost06,ost06a,ost10,ost10a}
has a serious drawback of 
neglecting energy-momentum correlations between multiple scattering
processes at the amplitude level~\cite{hla01}. Additionally,
the discussed treatment uses phenomenological parametrization for
Pomeron-Pomeron interaction vertices and neglects hard (high $|q^{2}|$)
$\Pom\Pom$ coupling. Hence, the scheme is unable to describe
the dynamical evolution of the saturation scale in hadron-hadron scattering.

\tit{Understanding the {}``ridge'' in proton-proton scattering at 7 TeV}

\auth{Klaus Werner\footnote{In collaboration with Iu.$\,$Karpenko and T.$\,$Pierog.} (Subatech, Nantes) }

The CMS collaboration published recently results~\cite{cms-ridge}
on two-particle correlations in $\Delta\eta$ and $\Delta\phi$, in
$pp$ scattering at 7 TeV. Most remarkable is the discovery of a ridge-like
structure around $\Delta\eta=0$, extended over many units in $\Delta\eta$,
referred to as {}``the ridge'', in high multiplicity $pp$ events.
A similar structure has been observed in heavy ion collisions at RHIC,
and there is little doubt that the phenomenon is related to the hydrodynamical
evolution of matter. This {}``fluid dynamical behavior'' is actually
considered to be the major discovery at RHIC.%

\begin{figure}[htpb]
\centering
\includegraphics[scale=0.45]{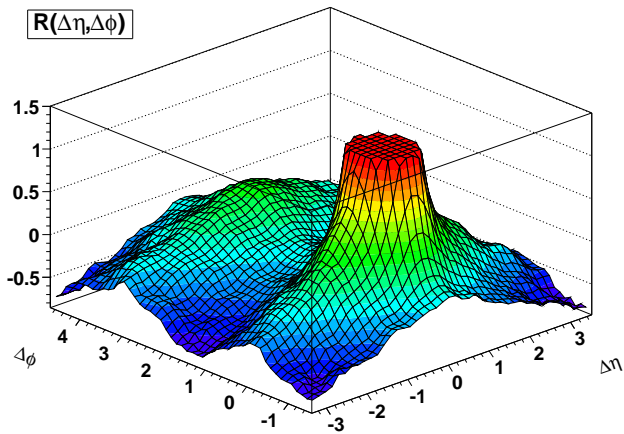}
 \vskip -0.2cm
\caption{Two-particle correlation function $R$ versus $\Delta\eta$
and $\Delta\phi$ for high multiplicity events in $pp$ collisions
at 7 TeV, as obtained from a hydrodynamical evolution based on flux
tube initial conditions. We consider particles with $p_T$ between
$1$ and 3 GeV/c. \label{cap:ridge} }
\end{figure}

So does $pp$ scattering provide as well a liquid, just ten times
smaller than a heavy ion collision? It seems so! We showed recently
\cite{epos2pp} that if we take exactly the same hydrodynamic approach
which has been so successful for heavy ion collisions at RHIC~\cite{epos2},
and apply it to $pp$ scattering, we obtain already very encouraging
results compared to $pp$ data at 0.9 TeV. In this paper, we apply
this fluid approach, always the same procedure, to understand the
7 TeV results. In Fig.~\ref{cap:ridge}, we show that our hydrodynamic
picture indeed leads to a near-side ridge, around $\Delta\phi=0$,
extended over many units in $\Delta\eta$. For the pure basic string
model, without hydro evolution, one finds no ridge! This shows that
the hydrodynamical evolution {}``makes'' the effect. 

It is easy to understand the origin of the ridge, in a hydrodynamical
approach based on flux tube initial conditions, see~\cite{epos_ridge}.
Imagine many (say 20) flux tubes of small transverse size (radius
$\approx0.2$ fm), but very long (many units of space-time rapidity
$\eta_{s}$ ). For a given event, their transverse positions are randomly
distributed within the overlap area of the two protons. Even for zero
impact parameter (which dominated for high multiplicity events), this
randomness produces azimuthal asymmetries. The energy density obtained
from the overlapping flux tubes shows an elliptical shape. And since
the flux tubes are long, and only the transverse positions are random,
we observe the same asymmetry at different longitudinal positions.
So we observe a translational invariant azimuthal asymmetry! 

If one takes this asymmetric but translational invariant energy density
as initial condition for a hydrodynamical evolution, the translational
invariance is conserved, and in particular translated into other quantities,
like the flow. At a later time, at different space-time rapidities,
the flow is more developed along the direction perpendicular to the
principal axis of the initial energy density ellipse. This is a 
typical fluid dynamical phenomenon, referred to as elliptic flow.
Important for this discussion: the asymmetry of the flow is again
translational invariant, the same for different values of~$\eta_{s}$.
Finally, particles are produced from the flowing liquid, with a preference
in the direction of large flow. This preferred direction is therefore
the same at different values of $\eta_{s}$. And since $\eta_{s}$
and pseudorapidity $\eta$ are highly correlated, one observes a $\Delta\eta$$\Delta\phi$
correlation, around $\Delta\phi=0$, extended over many units in $\Delta\eta$:
a particle emitted a some pseudorapidity $\eta$ has a large chance
to see a second particle at any pseudorapidity to be emitted in the
same azimuthal direction.

\tit{\fluka\ Monte Carlo}

\auth{Maria Vittoria Garzelli\footnote{G. Battistoni, M.V. Garzelli, A. Margiotta, S. Muraro, M. Sioli 
for the \fluka\ Collaboration} (INFN Milano \& Univ. Granada)}

The \fluka\ Monte Carlo code~\cite{mv1}, 
also interfaced with the \dpmjet\ code~\cite{mv2} 
for the treatment of nucleus-nucleus interactions, is being used in cosmic-ray
physics.  More detailed information on the physics
models for $h$-$h$, $h$-$A$, $A-A$ interactions relevant for 
cosmic-ray Physics adopted in \fluka\ and \dpmjet\ is available 
in the literature~\cite{mv3}.
Experimental observables involving muons, like cosmic 
$\mu^+/\mu^-$ charge ratios
and the $\mu$ decoherence function of $\mu$ bundles detected underground, 
can be used to test the hadronic interaction models.

\begin{figure}[htbp]
\centering 
\includegraphics[width=0.4\textwidth]{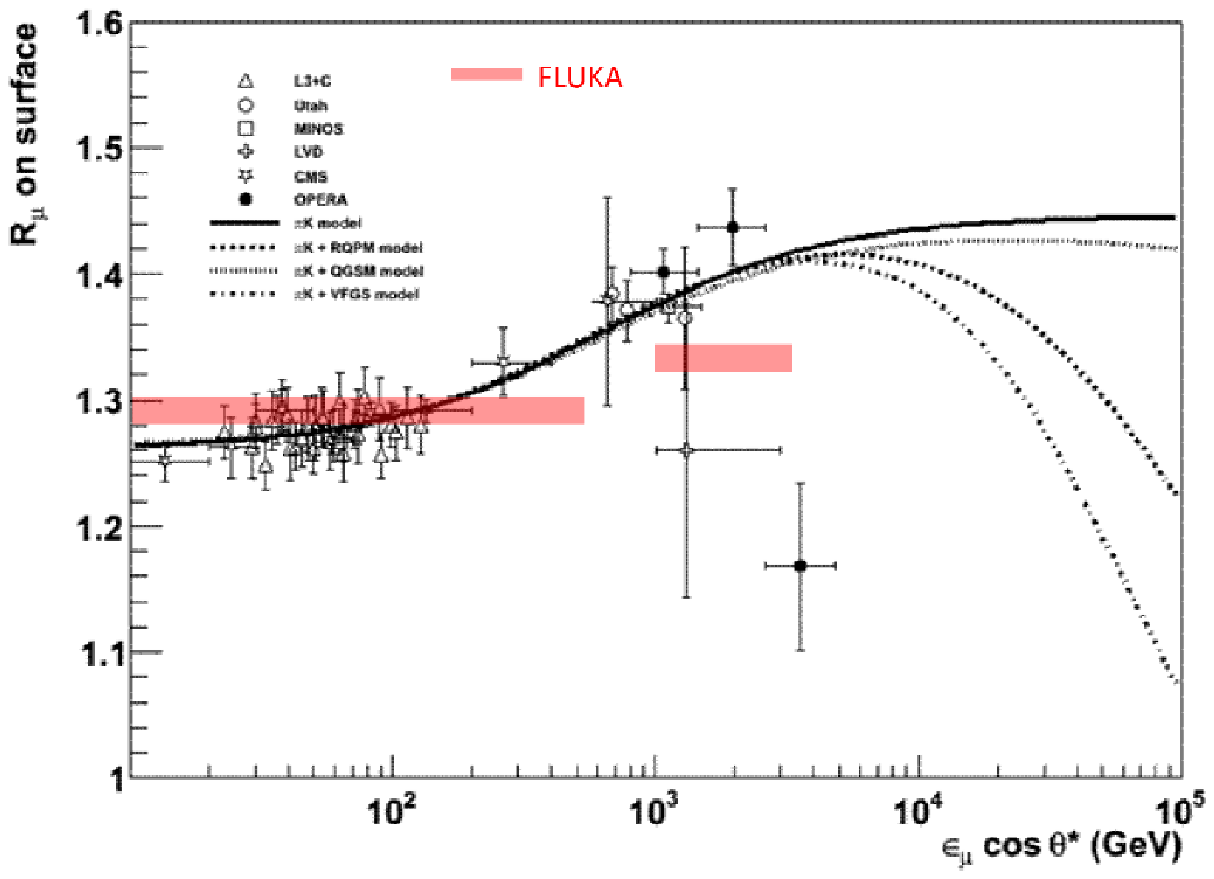}
\includegraphics[bb=1 1 521 381, width=0.4\textwidth]{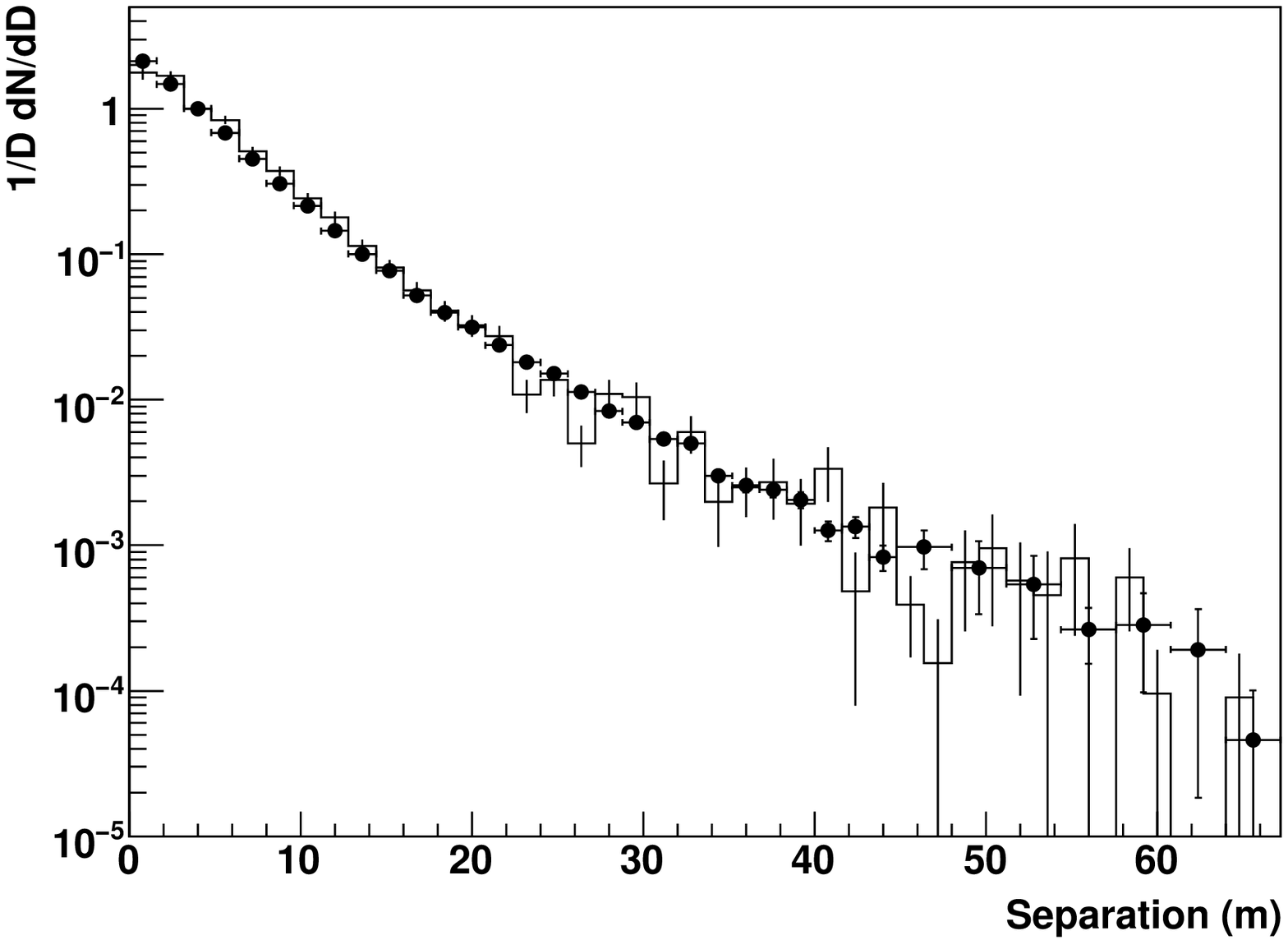}
\caption{\label{chargeratio} Left:
$\mu^+/\mu^-$ charge ratio as a function of the surface
energy $E_\mu {\mathrm cos}\theta$: 
\fluka\ simulation (pink bands representing
an average charge ratio over all $E_\mu {\mathrm cos}\theta$ 
values in the corresponding abscissa interval) 
vs. experimental data (Utah, LVD, L3+C, MINOS, CMS and OPERA). 
Right: decoherence distribution of $\mu$
pairs in a bundle (histogram: \fluka+\dpmjet; symbols: MACRO data)} 
\end{figure}

Cosmic $\mu^+/\mu^-$ charge ratios have been measured both by detectors at accelerator sites
(L3+C, CMS, preliminary data from ALICE) and by passive underground 
experiments (Utah, MINOS, OPERA). 
The results of \fluka\ simulations on  $\mu^+/\mu^-$ charge ratio are shown together
with available experimental data~\cite{mv4} in Fig.~\ref{chargeratio} (left). They 
turn out to be completely compatible with the CMS
and L3+C data, whereas they slightly underestimate the MINOS data (however not
completely confirmed by the OPERA ones, especially at the highest energies).
A possible reason of this discrepancy is the fact that the $K^+/K^-$  charge
ratio or the ratio between $K^+/K^-$  and  
$\pi^+/\pi^-$  charge ratios in \fluka\ can be underestimated. The
differences in the steepness of $x_{Feynman}$ distributions of $K^+$ and
$K^-$ is also fundamental.
To test these hypotheses new data on $K$ and $\pi$ production
at high-energy accelerators are urgently needed.

The decoherence function, i.e. the
distribution of the average distance between $\mu$ pairs in a bundle
detected underground, is another observable that allows one to test the
transverse structure predicted by the hadronic interaction models 
used to interpret the results of experiments on cosmic-rays. 
Such a measurement is highly sensitive to the $p_T$ distributions
of $\pi$ and $K$ produced in the first stages of the
shower formation process. We have verified that the decoherence function
is robust against a change of the details of the cosmic-ray primary
spectrum. \fluka\ + \dpmjet\ reproduces the experimental data from the MACRO
experiment~\cite{mv5}, as shown in Fig.~\ref{chargeratio} (right).

\tit{Comparison of model predictions to LHC data}

\auth{Tanguy Pierog (KIT, Karlsruhe)}

\begin{figure}[htbp]
\centering
\includegraphics[width=6.5cm,angle=-90]{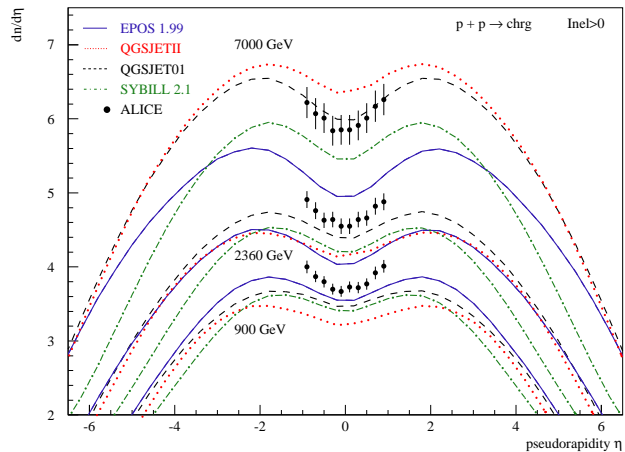}
\caption{Pseudorapidity distribution of charged particles measured by ALICE in $p p$ ($Inel>0$)
collisions compared to the predictions of \qgsjet\ 01 and II-03, 
 \sibyll\ 2.1, and \epos\ 1.99 models.}
\label{fig:eta}
\end{figure}

In April 2010, the ALICE~\cite{alice-jinst} collaboration published for the first time the 
pseudorapidity distribution of charged particles at a center-of-mass energy of 7~TeV~\cite{alice_mult2}. 
The analysis is based on a trigger called Inel$>$0 which consist of having at least one 
particle with $|\eta|<1$. Unlike the Non Single Diffractive trigger, this does not include 
any model-based correction and allows then for an easy comparison with any hadronic interaction 
models. In Fig.~\ref{fig:eta} we compare the ALICE data at 900~GeV, 2.36 and 7~TeV with the models
commonly used in air shower simulations, namely \qgsjet\ 01~\cite{qgsjet01} and II-03~\cite{qgsjetII}, 
\sibyll\ 2.1~\cite{Ahn:2009wx} and \epos\ 1.99~\cite{epos}, using the same trigger.

Even if none of these MCs can reproduce perfectly all the data, it is interesting to 
notice that the experimental results are well bracketed by the models used for cosmic ray analysis. 
The two extremes are \epos\ 1.99 which underestimate the growth of the particle density 
and \qgsjet\ II-03 which overestimates it. As a consequence we can say that the predictions 
of the Monte-Carlo models based on Regge field theory used for extended air shower analysis 
are compatible with the recent data from LHC, and hence do not show any tendency of dramatic 
change in minimum bias hadronic physics. Detailed results and dependence on trigger conditions 
not shown here indicate that soft physics is not yet completely understood.

\appendix

\clearpage

\section{List of participants}

\begin{wparticipants}
\parbox[t]{0.55\textwidth}{
\spk{Bruno Alessandro} {ALICE, Torino},\\ 
\spk{Doug Bergman} {HiRes, Salt Lake City},\\ 
\spk{Massimo Bongi} {LHCf, Florence},\\ 
\spk{Armen Bunyatyan} {HERA, Yerevan \& Hamburg},\\ 
\spk{Lorenzo Cazon} {Auger, Lisbon},\\
\spk{David d'Enterria} {Geneva},\\ 
\spk{Ivan de Mitri} {ARGO-YBJ, Lecce},\\ 
\spk{Paul Doll} {KASCADE-Grande, Karlsruhe},\\ 
\spk{Ralph Engel} {Karlsruhe},\\ 
\spk{Karsten Eggert} {TOTEM, Cleveland},\\ 
\spk{Maria Garzelli} {\fluka, Milano},\\
\spk{Lisa Gerhardt} {IceCube, Berkeley},\\ 
\spk{Stefan Gieseke} {\herwig, Karlsruhe},\\ 
\spk{Rohini Godbole} {Bangalore \& Geneva},\\ 
\spk{Jan Fiete Grosse-Oetringhaus} {ALICE, Geneva},\\ 
\spk{G\"osta Gustafson} {Lund},\\ 
\spk{Thomas Hebbeker} {CMS, Aachen},\\ 
\spk{Katsuaki Kasahara} {Tokyo},\\ 
\spk{Yoshio Kitadono} {Tsukuba},\\
}
\parbox[t]{0.45\textwidth}{
\spk{Lev Kheyn} {CMS, Moscow},\\ 
\spk{Joanna Kiryluk} {STAR, Berkeley},\\ 
\spk{Paolo Lipari} {Roma},\\ 
\spk{Sergey Ostapchenko} {\qgsjet, Tromso},\\ 
\spk{Tanguy Pierog} {\corsika, Karlsruhe},\\ 
\spk{Olga Piskounova} {{\sc qgsm}, Moscow},\\ 
\spk{Johannes Ranft} {\dpmjet-III, Siegen},\\ 
\spk{Amir Rezaeian} {Valparaiso},\\ 
\spk{Andrey Rostovtsev} {Moscow},\\ 
\spk{Nobuyuki Sakurai} {TA, Osaka},\\ 
\spk{Sebastian Sapeta} {Paris},\\ 
\spk{Sebastian Schleich} {LHCb, Dortmund},\\ 
\spk{Holger Schulz} {Berlin},\\ 
\spk{Torbj\"orn Sj\"ostrand} {\pythia\ 8, Lund},\\ 
\spk{Lars Sonnenschein} {Tevatron, Aachen},\\ 
\spk{Mark Sutton} {ATLAS, Sheffield},\\ 
\spk{Ralf Ulrich} {Karlsruhe},\\ 
\spk{Klaus Werner} {\epos\, Nantes},\\ 
\spk{Korinna Zapp} {\sherpa, Durham}
}
\end{wparticipants}

\vspace{-0.4cm}

\section{Programme}

\begin{tabbing}
xx:xx \= A very very very long name \= \kill

{\bf Monday, 29 November 2010}\\\vspace{0.7cm}
09:00 \hspace{0.4cm} \> {\it ECT* Director Welcome } (15')   \> \hspace{7.cm} A. Richter\\
09:15 \hspace{0.4cm} \> {\it Workshop introduction } (15')   \> \hspace{7.cm} D. d'Enterria\\
09:30 \hspace{0.4cm} \> {\it ATLAS QCD results } (30'+15')   \> \hspace{7.cm} M. Sutton\\
10:15 \hspace{0.4cm} \> {\it First CMS results } (30'+15')   \> \hspace{7.cm} T. Hebbeker\\
11:00 \hspace{0.4cm} \> {\it Particle multiplicity in p-p and HI collisions at colliders } (30'+15')   \> \hspace{7.cm} J.F. Grosse-Oetringhaus\\
14:00 \hspace{0.4cm} \> {\it LHCb QCD results } (30'+15')   \> \hspace{7.cm} S. Schleich\\
14:45 \hspace{0.4cm} \> {\it TOTEM results \& perspectives} (30'+15')   \> \hspace{7.cm} K. Eggert\\
15:30 \hspace{0.4cm} \> {\it LHCf results} (30'+15')   \> \hspace{7.cm} M. Bongi\\
16:00 \hspace{0.4cm} \> {\it Tevatron QCD results} (30'+15')   \> \hspace{7.cm} L. Sonnenschein \\
16:45 \hspace{0.4cm} \> {\it HERA results of relevance for cosmic rays} (30'+15')   \> \hspace{7.cm} A. Bunyatyan\\\\

{\bf  Tuesday 30 November 2010 }\\
09:00 \hspace{0.4cm} \> {\it Monte Carlo tuning at the LHC} (30'+15')   \> \hspace{7.cm}  H. Schulz\\
09:45 \hspace{0.4cm} \> {\it Parton Correlations and fluctuations} (30'+15')   \> \hspace{7.cm} G. Gustafson\\
10:30 \hspace{0.4cm} \> {\it \sherpa\ Monte Carlo} (30'+15')   \> \hspace{7.cm} K. Zapp\\
11:00 \hspace{0.4cm} \> {\it \herwig\ Monte Carlo} (30'+15')   \> \hspace{7.cm} S. Gieseke\\
11:45 \hspace{0.4cm} \> {\it \pythia\ 8} (30'+15')   \> \hspace{7.cm} T. Sj\"ostrand\\

14:00 \hspace{0.4cm} \> {\it Total cross sections at very high energies} (30'+15')   \> \hspace{7.cm} R. Godbole\\
14:45 \hspace{0.4cm} \> {\it Jets and the Underlying Event} (30'+15')   \> \hspace{7.cm} S. Sapeta\\
15:30 \hspace{0.4cm} \> {\it Gluon Saturation at the LHC from CGC} (30'+15')   \> \hspace{7.cm} A. Rezaeian\\\\

{\bf Wednesday 1 December 2010 }\\
09:00 \hspace{0.4cm} \> {\it ALICE cosmic ray results} (30'+15')   \> \hspace{7.cm} B. Alessandro\\
09:45 \hspace{0.4cm} \> {\it CMS results of relevance for cosmic-rays } (30'+15')   \> \hspace{7.cm} L. Kheyn\\
10:30 \hspace{0.4cm} \> {\it STAR results of relevance for cosmic rays } (30'+15')   \> \hspace{7.cm} J. Kiryluk \\
11:00 \hspace{0.4cm} \> {\it Hadron production in collider experiments } (30'+15')   \> \hspace{7.cm} A. Rostovtsev\\
11:45 \hspace{0.4cm} \> {\it Baryon production at collider energies} (30'+15')   \> \hspace{7.cm} O. Piskounova\\

14:00 \hspace{0.4cm} \> {\it \dpmjet-III: LHC predictions} (30'+15')   \> \hspace{7.cm} J. Ranft\\
14:45 \hspace{0.4cm} \> {\it Introduction to cosmic rays physis } (30'+15')   \> \hspace{7.cm} R. Engel\\
15:30 \hspace{0.4cm} \> {\it Cross section measurements using cosmic ray data } (30'+15') \> \hspace{7.cm} R. Ulrich\\\\

{\bf Thursday 2 December 2010 }\\
09:00 \hspace{0.4cm} \> {\it IceCube results} (30'+15')   \> \hspace{7.cm} L. Gerhardt\\
09:45 \hspace{0.4cm} \> {\it Telescope-Array results} (30'+15')   \> \hspace{7.cm} N. Sakurai\\
10:30 \hspace{0.4cm} \> {\it Auger results} (30'+15')   \> \hspace{7.cm} L. Cazon\\
11:00 \hspace{0.4cm} \> {\it ARGO-YBJ results} (30'+15')   \> \hspace{7.cm} I. De Mitri\\
11:45 \hspace{0.4cm} \> {\it KASCADE-Grande results} (30'+15')  \> \hspace{7.cm} P. Doll\\
14:00 \hspace{0.4cm} \> {\it HiRes results} (30'+15')  \> \hspace{7.cm} D. Bergman\\

14:45 \hspace{0.4cm} \> {\it UHECRs and hadronic interactions} (30'+15')   \> \hspace{7.cm} P. Lipari\\
15:30 \hspace{0.4cm} \> {\it Air shower predictions and features of hadronic interactions} (30'+15') \>\hspace{7.cm} R. Ulrich\\\\

{\bf Friday 3 December 2010 }\\
09:00 \hspace{0.4cm} \> {\it \sibyll\ Monte Carlo} (30'+15')   \> \hspace{7.cm} R. Engel\\
09:45 \hspace{0.4cm} \> {\it \qgsjet-II Monte Carlo} (30'+15')   \> \hspace{7.cm} S. Ostapchenko\\
10:30 \hspace{0.4cm} \> {\it \epos\ Monte Carlo} (30'+15')   \> \hspace{7.cm} K. Werner\\
11:00 \hspace{0.4cm} \> {\it \fluka\ Monte Carlo} (30'+15')   \> \hspace{7.cm} M. V. Garzelli \\
11:45 \hspace{0.4cm} \> {\it Comparison of UHCR models to LHC data} (30'+15')   \> \hspace{7.cm} T. Pierog \\\\

\end{tabbing}

%


\newpage



\begin{thebibliography}{10}

\bibitem{gzk}K.~Greisen, Phys. Rev. Lett. 16 (1966) 748--750; 
G.~T. Zatsepin and V.~A. Kuzmin, J. Exp. Theor. Phys. Lett. 4 (1966) 78.
\bibitem{Abbasi-2007-PRL-100-101101}R.~U. Abbasi {\it et~al.} [HiRes Collab.], Phys. Rev. Lett. 100, 101101 (2008) 
\bibitem{Abraham:2008ru}J.~Abraham {\it et al.}  [Pierre Auger Collab.], Phys.\ Rev.\ Lett.\  {\bf 101} (2008) 061101
\bibitem{d'Enterria:2007dt}
  D.~d'Enterria, Proceeds. DIS'07, Munich; arXiv:0708.0551 [hep-ex]. 


\bibitem{atlasone} 
ATLAS Collab., Phys. Lett. B 688 (2010) 21-42;
ATLAS-CONF-2010-024 (2010).

\bibitem{atlasthree}
ATLAS Collab., ATLAS-CONF-2010-046 (2010).

\bibitem{atlasfour}
ATLAS Collab., ATLAS-CONF-2010-029, ATLAS-CONF-2010-081 (2010).

\bibitem{atlasfive}
ATLAS Collab., accepted by EPJC, CERN-PH-EP-2010-034.

\bibitem{atlassix}
ATLAS Collab., ATLAS-CONF-2010-083 (2010).

\bibitem{atlasseven}
ATLAS Collab., ATLAS-CONF-2010-077 (2010).

\bibitem{atlaseight}
ATLAS Collab., submitted to JHEP, CERN-PH-EP-2010-037 (2010).

\bibitem{atlasnine} ATLAS Collab., ATLAS-CONF-2010-092 (2010). 



   
\bibitem{cms}
  CMS Coll., 
  JINST 3 (2008) S08004. 

\bibitem{cms-jets}
CMS Collab., 
CMS PAS QCD-10-011 (2010).

\bibitem{cms-dijets}
CMS Collab., 
Phys.Rev.Lett. 105 (2010) 211801.  

\bibitem{cms-rapidity}
CMS Collab., 
arXiv:1011.5531 [hep-ex].  

\bibitem{cms-strange}
CMS Collab., 
CMS PAS QCD-10-007 (2010).

\bibitem{cms-jetshapes}
CMS Collab., 
CMS PAS QCD-10-014 (2010).

\bibitem{cms-eventshapes}
CMS Collab., 
CMS PAS QCD-10-013 (2010).

\bibitem{cms-ridge}
CMS Collab., 
J. High Energy Phys. 9 (2010) 91. 

\bibitem{cms-talk}
CMS Collab., 
B. Wyslouch, ``Pb Pb collisions in CMS'', presentation CERN Dec 2, 2010.

\bibitem{cms-wz}
CMS Collab., 
CMS PAS EQK-10-002 (2010).

\bibitem{cms-top}
CMS Collab., 
arXiv:1010.5994 [hep-ex]. 


\bibitem{jfgo_review}J.F. Grosse-Oetringhaus, K. Reygers, 
J.\ Phys.\ G {\bf 37} (2010) 083001

\bibitem{alice_mult1}
  ALICE Collab., 
Eur.\ Phys.\ J.\  C {\bf 68} (2010) 89


\bibitem{alice_mult2}
  ALICE Collab., 
Eur.\ Phys.\ J.\  C {\bf 68} (2010) 345

\bibitem{cms_mult1}
  V.~Khachatryan {\it et al.}  [CMS Collab.],
  Phys.\ Rev.\ Lett.\  {\bf 105} (2010) 022002



\bibitem{LHCBDETECTOR}A. Augusto Alves Jr {\it et al.} [LHCb Collab.], JINST 3, S08005 (2008).
\bibitem{Aaij:2010nx}R.~Aaij {\it et al.}  [LHCb Collab.], Phys.\ Lett.\  B {\bf 693} (2010) 69. 
\bibitem{SkandsPerugia}  P.~Z.~Skands, Phys.\ Rev.\  D {\bf 82} (2010) 074018  


\bibitem{totem}See TOTEM web-page: \ttt{http://totem.web.cern.ch/Totem/}



\bibitem{UA7-1990}
E. Par\'e {\it et al.}, Phys. Lett. B, 242 (1990) 531--535

\bibitem{TDR-2006}
O. Adriani {\it et al.}, CERN-LHCC-2006-004, LHCF-TDR-001 (2006)

\bibitem{NIM578-2007}
T. Sako {\it et al.}, Nucl. Instr. and Meth. A 578 (2007) 146--159

\bibitem{ACTA38-2007}
R. D'Alessandro {\it et al.}, Acta Physica Polonica B 38 (2007) 829--838

\bibitem{JINST3-2008}
O. Adriani {\it et al.}, JINST 3 (2008) S08006

\bibitem{NIM612-2010}
M. Bongi {\it et al.}, Nucl. Instr. and Meth. A, 612 (2010) 451--454

\bibitem{JINST5-2010}
O. Adriani {\it et al.}, JINST 5 (2010) P01012

\bibitem{APH-2010}
H. Menjo {\it et al.}, Astroparticle Physics, In Press, 10.1016/j.astropartphys.2010.11.002


\bibitem{d0_det} D\O\ Collab., 
Nucl. Instrum. Methods Phys. Res., Sect. A {\bf 565}, 463 (2006). 


\bibitem{cdf_det} CDF Collab., 
Phys. Rev. D {\bf 71}, 032001 (2005) and refs. therein.

\bibitem{d0_6065_2010} D\O\ Collab.,  
D\O-Note 6065-Conf (2010).

\bibitem{cdf_prl102_2009} CDF Collab.,  
Phys. Rev. Lett. {\bf 102}, 222002 (2009). 

\bibitem{cdf_arXiv1007.5058_2010} CDF Collab., 
accepted at Phys. Rev. D, arXiv:1007.5048 (2010).

\bibitem{d0_6042_2010} D\O\ Collab., 
D\O\ Note 6042-Conf FERMILAB-PUB-10-361-E (2010).

\bibitem{cdf_prd77_2008} CDF Collab.,  
Phys. Rev. D {\bf 77}, 05204, hepex-0712.0604 (2008).

\bibitem{cdf_prelim_2006} CDF Collab., 
prelim. {\tt http://www-cdf.fnal.gov/physics/new/qcd/ QCD.html} (2006).


\bibitem{cdf_prl102a_2009} CDF Collab., 
Phys. Rev. Lett. {\bf 102}, 242001. 


\bibitem{cdf_prl_99_2007} CDF Collab., 
Phys. Rev. Lett. {\bf 99}, 242002 (2007).

\bibitem{cdf_prl_98_2007} CDF Collab., 
Phys. Rev. Lett. {\bf 98}, 112001 (2007).




\bibitem{d0_6054_2010} D\O\ Collab., 
D\O-Note 6054-Conf (2010).

\bibitem{cdf_10084_2010} CDF Collab., 
CDF/PUB/QCD/PUBLIC/10084 (2010).

\bibitem{d0_prd81_2010} D\O\ Collab., 
Phys. Rev. D {\bf 81}, 052012 (2010).  

\bibitem{cdf_prd82_2010} CDF Collab., 
Phys. Rev. D {\bf 82}, 034001, arXiv:1002.3146 (2010).

\bibitem{cdf_prelim_2010} CDF Collab., 
prelim. {\tt http://www-cdf.fnal.gov/physics/new/qcd/ QCD.html} (2010).

\bibitem{cdf_prl102b_2009} CDF Collab. 
Phys. Rev. Lett. {\bf 102}, 232002 (2009).



\bibitem{star-detector}  STAR Collab.,   K.H. Ackermann {\em{et al.}}, Nucl. Instrum. Meth. {\bf{A499}} (2003) 624.
\bibitem{pp} STAR Collab.,   B.I. Abelev  {\em{et al.}},  Phys. Rev. Lett. {\bf{97}} (2006) 252001; 
Phys. Rev. {\bf{D80}} (2009) 111108;  
J. Adams   {\em{et al.}},  Phys. Lett. {\bf{B616}} (2005) 8;  
{\em ibid.} Phys. Lett. {\bf{B637}} (2006) 161.
\bibitem{dAu-star}STAR Collab.,   J. Adams   {\em{et al.}},  Phys. Rev. Lett. {\bf{97}}  (2006) 152302. 
\bibitem{pp-w}
B. Surrow,  to appear in the SPIN2010 conference proceedings.
\bibitem{prompt}V.P. Goncalves and M.V.T. Machado,  JHEP 0704 (2007) 028 and refs. therein 
\bibitem{pp-non-photonic} STAR Collab.,  W. Xie {\em{et al.}},  POS DIS2010 (2010) 182;  
PHENIX Collab., A.Adare {\em{et al.}},  submitted to Phys. Rev. C; arXiv:1005.1627. 
\bibitem{fonll}M. Cacciari, P. Nason, and R. Vogt, Phys. Rev. Lett. {\bf{95}} (2005) 122001.
\bibitem{pp-J-Psi}STAR Collab.,  B.I. Abelev  {\em{et al.}},  Phys. Rev. {\bf{C80}} (2009) 041902. 
\bibitem{NRQCD} G. C. Nayak {\em{et al.}},  Phys. Rev. {\bf{D68}} (2003) 034003.
\bibitem{J-Psi-correlation}STAR Collab.,  Z. Tang {\em{et al.}},  
arXiv:1012.0233. 
\bibitem{pp-Upsilon}STAR Collab.,   B.I. Abelev  {\em{et al.}},  Phys.Rev. {\bf{D82}} (2010) 12004.
\bibitem{CEM} A.D. Frawley, T. Ullrich and R. Vogt, Phys. Rept. {\bf{462}} (2008) 125.
\bibitem{dAu-brahms}BRAHMS Collab.,  I. Arsene  {\em{et al.}}, Phys. Rev. Lett. {\bf{93}}  (2004) 242303.
\bibitem{di-hadron}STAR Collab.,  E. Braidot  {\em{et al.}},  arXiv:1005.2378. 
\bibitem{CGCsat} D.~Kharzeev, Y.V.~Kovchegov, and K.~Tuchin, Phys.Rev. {\bf{D68}} (2003) 094013; 
J.L. Albacete and C.Marquet, Phys. Rev. Lett. {\bf{105}} (2010) 162301 and refs. therein.
\bibitem{QGP} STAR Collab.,   J. Adams  {\em{et al.}},  Nucl. Phys. {\bf{A757}} (2005) 102;  
PHENIX Collab., K. Adcox  {\em{et al.}}, Nucl. Phys. {\bf{A757}} (2005) 184; 
PHOBOS Collab., B. B. Back {\em{et al.}},  Nucl. Phys. A757 (2005) 28;  
BRAHMS Collab., I. Arsene  {\em{et al.}},  Nucl. Phys. {\bf{A757}} (2005) 1. 
\bibitem{BES}STAR Collab.,  M.M. Aggarwal  {\em{et al.}},  arXiv:1007.2613 
\bibitem{anti-matter}STAR Collab.,   B.I. Abelev  {\em{et al.}}, Science {\bf{328}} (2010) 58. 



\bibitem{GG1}     E. Avsar, G. Gustafson, and L. L\"{o}nnblad,
     \emph{JHEP} {\bf 07} (2005) 062, \emph{ibid}  {\bf 01} (2007) 012,
    \emph{ibid}  \emph{JHEP} {\bf 12} (2007) 012.

\bibitem{GG2}      C. Flensburg, G. Gustafson, and L. L\"{o}nnblad,
     \emph{Eur. Phys. J.} {\bf C60} (2009) 233.

  C.~Flensburg and G.~Gustafson,
  JHEP {\bf 1010} (2010) 014. 
  


\bibitem{Ryskin:2009tj}
  M.~G.~Ryskin, A.~D.~Martin and V.~A.~Khoze,
  Eur.\ Phys.\ J.\  C {\bf 60} (2009) 249




\bibitem{Bahr:2008pv}
  M.~Bahr {\it et al.},
  Eur.\ Phys.\ J.\  C {\bf 58} (2008) 639

\bibitem{Bahr:2007ni}
  M.~Bahr {\it et al.},
  arXiv:0711.3137 [hep-ph].

\bibitem{Bahr:2008tx}
  M.~Bahr {\it et al.},
  arXiv:0804.3053 [hep-ph].

\bibitem{Bahr:2008tf}
  M.~Bahr {\it et al.},
  arXiv:0812.0529 [hep-ph].


\bibitem{Sjostrand:1987su}
  T.~Sj\"ostrand and M.~van Zijl,
  Phys.\ Rev.\  D {\bf 36} (1987) 2019.

\bibitem{Butterworth:1996zw}
  J.~M.~Butterworth, J.~R.~Forshaw and M.~H.~Seymour,
  Z.\ Phys.\  C {\bf 72} (1996) 637

\bibitem{Borozan:2002fk}
  I.~Borozan and M.~H.~Seymour,
  JHEP {\bf 0209}, 015 (2002)

\bibitem{Bahr:2008dy}
  M.~Bahr, S.~Gieseke and M.~H.~Seymour,
  JHEP {\bf 0807} (2008) 076

\bibitem{Bahr:2008wt}
  M.~Bahr, S.~Gieseke and M.~H.~Seymour,
  arXiv:0809.2669 [hep-ph].

\bibitem{Bahr:2008wk}
  M.~Bahr, J.~M.~Butterworth and M.~H.~Seymour,
  JHEP {\bf 0901} (2009) 065

\bibitem{Bartalini:2010su}
  P.~Bartalini {\it et al.},
  arXiv:1003.4220 [hep-ex].

\bibitem{Bahr:2009ek}
  M.~Bahr, J.~M.~Butterworth, S.~Gieseke and M.~H.~Seymour,
  arXiv:0905.4671 [hep-ph].

\bibitem{christian}
C.~R\"ohr, Diploma thesis, Karlsruhe Institute of Technology 2010


\bibitem{Affolder:2001xt}
  A.~A.~Affolder {\it et al.}  [CDF Collab.],
  Phys.\ Rev.\  D {\bf 65} (2002) 092002.

\bibitem{Acosta:2004wqa}
  D.~E.~Acosta {\it et al.}  [CDF Collab.],
  Phys.\ Rev.\  D {\bf 70} (2004) 072002




\bibitem{nchgeq6} ATLAS-CONF-2010-031


\bibitem{TS1} T. Sj\"ostrand, S. Mrenna and P. Skands, Comput. Phys. Comm. 178 (2008) 852
\bibitem{TS2} T. Sj\"ostrand, S. Mrenna and P. Skands, JHEP05 (2006) 026
\bibitem{TS3} R. Corke and T. Sj\"ostrand, Eur. Phys. J. C69 (2010) 1
\bibitem{TS4} R. Corke and T. Sj\"ostrand, arXiv:1011.1759 [hep-ph]
\bibitem{TS5} R. Corke and T. Sj\"ostrand, JHEP 01 (2010) 035


\bibitem{Godbole:2004kx}
 R.~M.~Godbole, A.~Grau, G.~Pancheri and Y.~N.~Srivastava,
 Phys.\ Rev.\  D {\bf 72}, 076001 (2005).

\bibitem{Godbole:2008ex}
  R.~M.~Godbole, A.~Grau, G.~Pancheri and Y.~N.~Srivastava,
  Eur.\ Phys.\ J.\  C {\bf 63}, 69 (2009)

\bibitem{corsetti} 
A. Corsetti, A. Grau, G. Pancheri,Y.N. Srivastava, Phys.\ Lett.\ {\bf B 382}, 282 (1996)

\bibitem{cornet}
F. Cornet, C.A. Garcia Canal, A. Grau, G. Pancheri and G. Sciutto,
Proceedings of the  31st ICRC, Lodz 2009.

\bibitem{lianew}
A. Grau, G. Pancheri,O. Shekhovtsova and Y. N. Srivastava, Phys.\ Lett.\  {\bf B 693}, 456 (2009).


  \bibitem{Cacciari:2009dp}
    M.~Cacciari, G.~P.~Salam and S.~Sapeta,
    JHEP {\bf 1004} (2010) 065

  \bibitem{KarFieldDY} 
    D.~Kar and R.~Field (CDF Collab.),
    CDF/PUB/CDF/PUBLIC/9531, July (2008)

  \bibitem{Cacciari:2007fd}
    M.~Cacciari and G.~P.~Salam,
    Phys.\ Lett.\  B {\bf 659} (2008) 119

  \bibitem{Cacciari:2008gn}
    M.~Cacciari, G.~P.~Salam and G.~Soyez,
    JHEP {\bf 0804} (2008) 005

  \bibitem{CMS-UE} 
    CMS Collab., 
    CMS-PAS-QCD-10-005, July (2010)


\bibitem{CGC}See e.g. F. Gelis {\it et al.}, 
arXiv:1002.0333; L. McLerran, arXiv:1011.3203, arXiv:1011.3204. 

\bibitem{LRPP} E.~Levin and A.~H.~Rezaeian, Phys.\ Rev.\  {\bf D82 } (2010)  014022; 
and arXiv:1011.3591. 


\bibitem{rid}A. Dumitru {\it et al.}, 
arXiv:1009.5295.
\bibitem{alice}  K.~Aamodt {\it et al.}  [ALICE Collab.],
  arXiv:1011.3916. 


\bibitem{talkme}Talk given by A. H. Rezaeian in this workshop. 

\bibitem{LR2}E.~Levin and A.~H.~Rezaeian, Phys.\ Rev.\  {\bf D82 } (2010)  054003. 


\bibitem{BR}A. Bylinkin, A. Rostovtsev, e-Print: arXiv:1008.0332 [hep-ph].



\bibitem{ua1} UA1 Collab., G. Bocquet {\it et.al.}, Z. Phys. C {\bf 366} (1996) 441.
\bibitem{cdf} CDF Collab., D. Acosta {\it et.al.}, Phys. Rev. D {\bf 72} (2005) 052001.
\bibitem{star} STAR Collab., B.I. Abelev {\it et.al.}, Phys. Rev. C {\bf 75} (2007) 064901.
\bibitem{isr} ISR Collab., D.Drijard {\it et.al.}, Z.Phys. C {\bf 12} (1982) 217.
\bibitem{wa89} WA89 Collab., M.I.Adamovich {\it et.al.}, Eur. Phys. J. C {\bf 26}(2003) 357.
\bibitem{qgsm} A.B. Kaidalov and K.A. Ter-Martirosyan, Sov. J. Nucl. Phys. {\bf 39}, 1545 (1984); 
{\bf 40},211(1984); A.B. Kaidalov, Phys. Lett. B {\bf 116}, 459 (1982).


\bibitem{phojet-a}
R.\ Engel:
Z. Phys. C \textbf{66}, 203 (1995),
R.\ Engel and J.\  Ranft:
Phys. Rev. D \textbf{54}, 4244 (1996)

\bibitem{dpmjet2}
S.\ Roesler, R.\ Engel and J.\  Ranft:
hep--ph/0012252, Proc. of Monte Carlo 2000, Lisboa, Oct.2000,
Springer,p.1033



\bibitem{Auger} Pierre Auger Collab., Nucl. Instruments and Methods A523 (2004), 50.
\bibitem{SDFDHY} Pierre Auger Collab., Nucl. Instruments and Methods in Phys. Research A613 (2010), 29-39; A620 (2010) 227; and arXiv:1010.6162 
\bibitem{spectrum} Pierre Auger Collab., Phys. Lett. B685 (2010) 239.
\bibitem{Xmax} Pierre Auger Collab., Phys. Rev. Lett. 104 (2010) 091101. 
\bibitem{muons} Pierre Auger Collab., Proceeds. 31st ICRC in Lodz, Poland (2009); arXiv:0906.2319. 
\bibitem{neutrinos} Pierre Auger Collab., Phys. Rev.  D79  (2009), 102001.
\bibitem{photons} Pierre Auger Collab., Astropart. Phys.  31  (2009), 399.
\bibitem{science} Pierre Auger Collab., Science 318 (2007), 939; Astropart. Phys. 29 (2008), 188.
\bibitem{UpdateAGN} Pierre Auger Collab., Astropart. Phys. 34 (2010) 314. 




\bibitem{Abbasi-2009-APP-32-53}R.~U. Abbasi {\it et al.} [HiRes Collab.], Astropart. Phys. 32, 53 (2009) 

\bibitem{Berezinsky-Grigoreva-1988-AA-199-1}V.~S. Berezinsky and S.~I. Grigor'eva, Ast. and Astrophys. 199, 1 (1988).

\bibitem{Abbasi-2010-PRL-104-161101}R.~U. Abbasi {\it et al.} [HiRes Collab.], Phys. Rev. Lett. 104, 161101 (2010) 

\bibitem{Abbasi-2008-APP-30-175}R.~U. Abbasi {\it et al.} [HiRes Collab.], Astropart. Phys. 30, 175 (2008) 

\bibitem{Abbasi-2010-arXiv:1002.1444}R.~U. Abbasi {\it et al.} [HiRes Collab.], arXiv:1002.1444.



\bibitem{bacci2002}
C. Bacci {\it et al.}, Astropart. Phys. {\bf 17} (2002) 151 and references therein
\bibitem{demitri2007cris}
I. De Mitri {\it et al.}, Nucl. Phys. {\bf B} (Proc. Suppl.) {\bf 165} (2007) 66

\bibitem{xsecpaper}
G.Aielli {\it et al.}, Phys. Rev. D {\bf 80} 092004 (2009) and references therein

\bibitem{ulrich2007}
R.Ulrich {\it et al.}, New J. Phys. {\bf 11} (2009) 065018

\bibitem{production}
R. Engel {\it et al.}, Phys. Rev. D {\bf 58} (1988) 014019 

\bibitem{pairtopp}
See e.g. T.K. Gaisser, Phys. Rev D {\bf 36} (1987) 1350 and M.M. Block, Phys. Rev D {\bf 76} (2007) 111503 and refs. therein 

\bibitem{blockhalzen2005}
M.M. Block and F.Halzen, Phys. Rev. {\bf D72} (2005) 036006


\bibitem{PD1}D. Kang {\it et al.}, arXiv:1009.4902 [astro-ph.HE].
\bibitem{PD2}P. Doll {\it et al.}, 
 arXiv:1010.2702 [astro-ph.HE].  


\def\etal{{\it et al.}}
\bibitem{pmt}C. Wiebusch \etal, Nucl. Instr. Meth. Phys. Res. A {\bf 618} 139 (2010).
\bibitem{ps1}R. Abbasi \etal\ [HiRes Collab.], Astrophys. Jour. Lett. {\bf 701} L47 (2009).
\bibitem{ps2}R. Abbasi \etal\ [HiRes Collab.], Phys. Rev. Lett. {\bf 103}, 221102 (2009).
\bibitem{ehe}R. Abbasi \etal\ [HiRes Collab.], Phys. Rev. D {\bf 82}, 072003 (2010).
\bibitem{grb}R. Abbasi \etal\ [HiRes Collab.], Astrophys. Jour. {\bf 710}, 346 (2010).
\bibitem{wimp1} R. Abbasi \etal\ [HiRes Collab.], Phys. Rev. Lett. {\bf 102}, 201302 (2010).
\bibitem{wimp2}R. Abbasi \etal\ [HiRes Collab.], Phys. Rev. D {\bf 81}, 057101 (2010).
\bibitem{atm}R. Abbasi \etal\ [HiRes Collab.], submitted to Phys. Rev. D (2010).
\bibitem{aniso}R. Abbasi \etal\ [HiRes Collab.], Astrophys. Jour. {\bf 718} L194 (2010).


\bibitem{Bluemer:2009zf}
J.~Bl{\"u}mer, R.~Engel, and J.~R. H{\"o}randel,
Prog. Part. Nucl. Phys. 63 (2009) 293--338

\bibitem{Stanev:1993tx}
T.~Stanev, P.~L. Biermann, and T.~K. Gaisser,
Astron. \& Astroph. 274 (1993) 902

\bibitem{Erlykin:2010yy}
A.~Erlykin, T.~Wibig, and A.~W. Wolfendale,
arXiv:1009.0600 

\bibitem{Hoerandel:2002yg}
J.~R. H{\"o}randel,
Astropart. Phys. 19 (2003) 193--220

\bibitem{Hillas:2005cs}
A.~M. Hillas,
J. Phys. G31 (2005) R95--R131.

\bibitem{Berezinsky:2002nc}
V.~Berezinsky, A.~Z. Gazizov, and S.~I. Grigorieva,
Phys. Rev. D74 (2006) 043005

\bibitem{Knapp:2002vs}
J.~Knapp, D.~Heck, S.~J. Sciutto, M.~T. Dova, and M.~Risse,
Astropart. Phys. 19 (2003) 77--99




\bibitem{Abraham:2004dt}
J.~Abraham {\it et~al.}  (Pierre Auger Collab.),
Nucl. Instrum. Meth. A523 (2004) 50.

\bibitem{Abbasi:2004nz}
R.~Abbasi {\it et~al.}  (HiRes Collab.),
Astrophys. J. 622 (2005) 910--926

\bibitem{Kawai:2008zza}
H.~Kawai {\it et~al.}  (TA Collab.),
Nucl. Phys. Proc. Suppl. 175-176 (2008) 221--226.

\bibitem{Ulrich:2010rg}
R.~Ulrich, R.~Engel, and M.~Unger,
arXiv:1010.4310 

\bibitem{Abraham:2010yv}
J.~Abraham {\it et~al.}  (Pierre Auger Collab.),
Phys. Rev. Lett. 104 (2010) 091101

\bibitem{Abbasi:2009nf}
R.~U. Abbasi {\it et~al.}  (HiRes Collab.),
Phys. Rev. Lett. 104 (2010) 161101


\bibitem{Fletcher:1994bd}
R.~S. Fletcher, T.~K. Gaisser, P.~Lipari, and T.~Stanev,
Phys. Rev. D50 (1994) 5710--5731.


\bibitem{Ahn:2009wx}
E.-J. Ahn, R.~Engel, T.~K. Gaisser, P.~Lipari, and T.~Stanev,
Phys. Rev. D 80 (2009) 094003

\bibitem{Good:1960ba}
M.~L. Good and W.~D. Walker,
Phys. Rev. 120 (1960) 1857--1860.

\bibitem{Gluck:1994uf}
M.~Gl{\"u}ck, E.~Reya, and A.~Vogt,
Z. Phys. C67 (1995) 433--448.

\bibitem{Gribov:1984tu}
L.~V. Gribov, E.~M. Levin, and M.~G. Ryskin,
Phys. Rept. 100 (1983) 1--150.

\bibitem{Alt:2005zq}
C.~Alt {\it et~al.}  (NA49 Collab.),
Eur. Phys. J. C45 (2006) 343--381;
{\it ibid} C49 (2007) 897--917


\bibitem{Glauber:1970jm}
R.~J. Glauber and G.~Matthiae,
Nucl. Phys. B21 (1970) 135--157.

\bibitem{Engel:1992vf}
J.~Engel, T.~K. Gaisser, T.~Stanev, and P.~Lipari,
Phys. Rev. D46 (1992) 5013--5025.


\bibitem{gri68} V.~N. Gribov,  Sov.~Phys.~JETP~{\bf 26}, 414 (1968); {\em ibid.}~{\bf 29}, 483 (1969).
\bibitem{agk}V.~A.~Abramovskii, V.~N.~Gribov and O.~V.~Kancheli, Sov.~J.~Nucl.~Phys.~{\bf 18},  308 (1974).
\bibitem{ost06}S.~Ostapchenko,   Phys.~Lett.~B~{\bf 636}, 40 (2006);  Phys.~Rev.~D~{\bf 77}, 034009 (2008).
\bibitem{ost06a}S.~Ostapchenko,     Phys.~Rev.~D~{\bf 74}, 014026 (2006).
\bibitem{ost10}S.~Ostapchenko,     Phys.~Rev.~D~{\bf 81}, 114028  (2010).
\bibitem{ost10a}S.~Ostapchenko, Phys.~Rev.~D~in press; arXiv:1010.1869. 
\bibitem{hla01}M.~Hladik  {\em et al.}, Phys.~Rev. Lett.~{\bf 86}, 3506 (2001).




\bibitem{epos2pp}K. Werner, Iu.$\,$Karpenko, T.$\,$Pierog, M. Bleicher,
K. Mikhailov, arXiv:1010.0400, submitted

\bibitem{epos2}K.$\,$Werner, Iu.$\,$Karpenko, T.$\,$Pierog, M.
Bleicher, K. Mikhailov, arXiv:1004.0805, to be published in Phys.
Rev. C

\bibitem{epos_ridge}K. Werner, Iu.$\,$Karpenko, T.$\,$Pierog, arXiv:1011.0375










\bibitem{mv1}A. Fass\`o {\it et al.}, CERN Yellow Report 2005-10, INFN/TC\_05/11 (2005) 1; 
hep-ph/0306267. {\texttt http://www.fluka.org} 

\bibitem{mv2}S. Roesler {\it et al.}, hep-ph/0012252 and references therein. 
J. Ranft, Phys. Rev. D {\bf 51} (1995) 64.

\bibitem{mv3}G. Battistoni {\it et al.}, Nucl. Phys. Proc. Suppl. {\bf 175-176} (2008) 88; 
0711.2044; AIP Conf. Proc. {\bf 972} (2008) 449; 
arXiv:1002.4655. 

\bibitem{mv4}G.K. Ashley {\it et al.}, Phys. Rev. D {\bf 12} (1975) 20. 
L3 Collab., Phys. Lett. B {\bf 598} (2004) 15. 
MINOS Collab., Phys. Rev. D {\bf 76} (2007) 052003.
CMS Collab., Phys. Lett. B {\bf 692} (2010) 83. 
OPERA Collab., EPJC {\bf 67} (2010) 25.  

\bibitem{mv5}MACRO Collab. hep-ex/9901027, Phys. Rev. D {\bf 60} (1999) 032001. 



\bibitem{alice-jinst}K.~Aamodt, {\it et al.}, [ALICE Collab.], JINST 3,S08002 (2008)
\bibitem{qgsjet01} N.N.~Kalmykov, S.S.~Ostapchenko and A.I.~Pavlov, Nucl. Phys. Proc. Supl. 52B, 17  (1997).
\bibitem{qgsjetII} S.\ Ostapchenko,  Phys.\ Rev.\  D {\bf 74}, 014026 (2006); 
 AIP Conf.\ Proc.\ {\bf 928}, 118  (2007).
\bibitem{epos} K. Werner {\it et al.}, Phys Rev. C 74 044902 (2006).


\end{thebibliography}
\end{document}